\documentclass[10pt,aps,prd,twocolumn,showpacs,superscriptaddress,nofootinbib,nobibnotes,longbibliography,floatfix]{revtex4-1}

\usepackage[utf8]{inputenc}
\usepackage[T1]{fontenc}
\usepackage{enumitem}
\usepackage{bm}
\usepackage{mathtools,amsmath,amssymb,amsfonts,mathrsfs,eucal,graphicx,tensor,csquotes,accents, commath,chngcntr,siunitx}
\usepackage[dvipsnames]{xcolor}
\usepackage[unicode]{hyperref}
\hypersetup{colorlinks=true, citecolor=MidnightBlue,
            linkcolor=MidnightBlue, urlcolor=MidnightBlue, linktocpage=true}
\usepackage[normalem]{ulem}
\usepackage{amsmath}
\usepackage{tensor}
\usepackage{leftindex}
\usepackage{amssymb} 

\usepackage[compat=1.1.0]{tikz-feynman}
\usepackage{pifont}

\newcommand{\tick}{\ding{51}}  
\newcommand{\cross}{\ding{55}} 

\DeclareMathAlphabet{\mathpzc}{OT1}{pzc}{m}{it}
\definecolor{darkgreen}{rgb}{0.0, 0.6, 0.0}

\newcommand{\g}{\mathsf{g}}

\tikzfeynmanset{warn luatex=false}
\begin{document}

\title{Perturbations of the Vaidya metric in the frequency domain: \\Quasi-normal modes and tidal response}

\author{Lodovico Capuano}
\affiliation{SISSA, Via Bonomea 265, 34136 Trieste, Italy and INFN Sezione di Trieste}
\affiliation{IFPU - Institute for Fundamental Physics of the Universe, Via Beirut 2, 34014 Trieste, Italy}

\author{Luca Santoni}
\affiliation{Universit\'e Paris Cit\'e, CNRS, Astroparticule et Cosmologie, 10 Rue Alice Domon et L\'eonie Duquet, F-75013 Paris, France}

\author{Enrico Barausse}
\affiliation{SISSA, Via Bonomea 265, 34136 Trieste, Italy and INFN Sezione di Trieste}
\affiliation{IFPU - Institute for Fundamental Physics of the Universe, Via Beirut 2, 34014 Trieste, Italy}

\begin{abstract}
The mass of a black hole can dynamically evolve due to various physical processes, such as for instance accretion, Hawking radiation, absorption
of gravitational/electromagnetic waves, superradiance, etc. This evolution can have an impact on astrophysical observables, like the  ringdown gravitational signal. An effective description of a spherically symmetric  black hole with evolving mass is provided by the Vaidya metric. In our investigation, we explore the dynamics of linear perturbations on this background, assuming a 
constant rate of change for the mass. Despite the time-dependent background, a judicious change of coordinates allows us to treat the perturbations in the frequency domain, and to compute explicitly the quasi-normal modes and the tidal Love numbers.
\end{abstract}

\maketitle
\section{Introduction}
The dynamical evolution of a binary system of black holes (BHs), and the ensuing emission of gravitational waves (GW), is conventionally divided
 in three distinct phases:  inspiral,  merger, and ringdown. During the initial inspiral phase, the binary progressively loses orbital energy and angular momentum through emission of gravitational radiation, leading to a spiraling motion of the objects towards each other, which  ends with the merger. After the merger, the remnant object undergoes damped oscillations known as quasi-normal modes (QNMs), as it relaxes to a final stable configuration  \cite{1970Natur.227..936V,89c00a43-362f-3c19-9cc2-fa0808f91935,Anninos_1993,Kokkotas_1999,Berti_2009,Konoplya_2011}. This last phase is called ringdown. 
GWs produced during the ringdown are generally modelled as perturbations on top of a 
stationary  BH background metric, without accounting for possible evolution of the BH mass. 
However, realistic BHs are generally non-stationary, as they can dynamically evolve due to the interplay with the surrounding environment. Modelling a dynamical BH background is then crucial for investigating diverse physical scenarios.

Astrophysical BHs are often surrounded by accretion disks, which lead to an increase in the BH mass~\cite{Pringle1981AccretionDI,Shakura1973BlackHI}. 
It is also known that BHs slowly evaporate through the Hawking mechanism \cite{1974Natur.248...30H, Hawking1975ParticleCB,PhysRevD.13.198}. The timescale of this process grows with the mass of the evaporating BH, hence its effect is likely irrelevant for stellar-origin and supermassive BHs, while it could in principle play a role in the dynamics of primordial ones. 
Another process that is capable of removing mass from a BH is given by superradiant instabilities driven by massive boson fields \cite{Brito_2020,HERDEIRO2022136835}. The unstable production of such particles can in fact be triggered around a BH at the expense of its energy and angular momentum.  
This effect is related to the very well-known Penrose process \cite{1971NPhS..229..177P}, and it requires the presence of an ergoregion. 

Note that while allowing the extraction of mass/energy from a stationary BH, Hawking evaporation and the Penrose process 
still satisfy the laws of BH thermodynamics \cite{Bardeen1973TheFL,PhysRevD.7.2333}. Moreover, one can consider more exotic scenarios, like accretion of a phantom matter field (i.e., a field violating the dominant energy condition, where BHs can actually lose mass while accreting \cite{deFreitasPacheco:2007ysm, Lima:2008qx}. Finally, 
the BH mass is expected to evolve during the ringdown phase also due to the absorption of ringdown GWs themselves \cite{Redondo-Yuste:2023ipg,Zhu:2024dyl}.  
This absorption of GWs is also expected to impact the inspiral dynamics in binary systems in the form of tidal heating \cite{Alvi_2001,Poisson_2004,Chatziioannou_2013,Isoyama_2017, Datta_2020,Saketh_2023}. 

Furthermore, in the inspiral phase, the GW signal is  affected by the tidal response of the binary objects. The conservative part of the response is generally quantified in terms of tidal Love numbers \cite{chakrabarti2013new, Cardoso:2017cfl, Franzin:2017mtq}, although more general parameters capturing dissipative effects can also be defined \cite{Chia_2021,Goldberger:2020fot, Saketh:2023bul}. Static tidal Love numbers are  zero for stationary, asymptotically-flat BHs in four-dimensional General Relativity (GR) \cite{Fang:2005qq,Damour:2009vw,Binnington:2009bb,Gurlebeck:2015xpa,Hui:2020xxx,Kol_2012,PhysRevD.103.084021,Charalambous_2021,Rodriguez:2023xjd} (even at nonlinear level for Schwarzschild BHs \cite{Gurlebeck:2015xpa,Poisson:2020vap,Riva:2023rcm}) suggesting the existence of hidden symmetries \cite{Hui:2021vcv,Hui:2022vbh,Charalambous:2021kcz,Charalambous:2022rre,Rai:2024lho}, while they can be be non-vanishing if one considers nontrivial asymptotics \cite{Nair:2024mya}, or more general theories of gravity \cite{Katagiri:2023umb,PhysRevD.108.084049}. 

A simple model of spherical, uncharged, dynamical BH spacetimes is provided by the Vaidya metric  (see \cite{Vaidya:1951fdr,RelativistToolkit} for a detailed introduction).\footnote{The electrically charged generalization was first introduced in \cite{Bonnor1970SphericallySR}, while rotating Vaidya solutions have been proposed in \cite{PhysRevD.1.3220, CARMELI197797}.} This model, while  relatively simple, provides an effective description of the processes mentioned above,  allowing  for an estimate of their impact on the astrophysical observables, e.g.~the GW signals. A formal discussion of the perturbative dynamics of the Vaidya background can be found in \cite{Mbonye_2000, Nolan_2005, Nolan_2006}. Studies of QNMs for the Vaidya metric have already been carried out in the time domain with a fully numerical approach  
in \cite{Shao:2004ws, Abdalla:2006vb,Chirenti:2010iu, Chirenti:2011rc, Lin:2021fth, Redondo-Yuste:2023ipg}.

In this work, we perform instead a semi-analytic calculation in the frequency domain based on the continued-fraction method (also known as Leaver's method) \cite{Leaver:1985ax}, which is the most accurate
technique for computing QNM frequencies, and  
is  capable of handling overtones. Our approach relies on the assumption that the rate of change of the mass is constant. In this way, a proper choice of the coordinates allows to factor out the time dependence in a conformal factor and perform Fourier transforms, as it is usually done in standard BH perturbation theory. 
In this framework, we also discuss the tidal response of a dynamical BH, arguing that  non-vanishing corrections to the Love numbers appear, as a result of the nontrivial mass evolution. 

The paper is organized as follows. In section \ref{Vaidya_Generalities}, we describe the mathematical properties of the Vaidya solution and its conformal structure in the case in which the mass derivative is constant. In section \ref{pert_eqs_Vaidya}, we derive the perturbation equations for scalar, electromagnetic and (axial) gravitational  fields on such background. In section \ref{QNMs}, we estimate the QNMs analytically in the eikonal limit, and numerically using Leaver's continued-fraction method. Finally, in section \ref{LoveNumbers}, we discuss the tidal response of the Vaidya BH. In particular, we will perform the matching with the point-particle effective theory, and show how a small mass derivative provides nontrivial perturbative corrections to the vanishing Schwarzschild Love number couplings. 

We will work in geometric units $c = G = 1$, and in the  mostly plus signature of the metric $(-,+,+,+)$.

\section{The causal structure of the Vaidya black hole}
\label{Vaidya_Generalities}

The Vaidya solution describes a (non-empty) spherically symmetric BH spacetime with dynamical mass. 
A compact way to write its line element is by taking the Schwarzschild line element in ingoing/outgoing Eddington--Finkelstein coordinates $(w,r,\theta,\phi)$ and by promoting the Schwarzschild mass parameter to be a function of the null coordinate $w$. 
The latter  can either be  the advanced time, $w = t+r_*$, or the retarded time, $w=t-r_*$, where $r_*$ is the  Schwarzschild tortoise coordinate and $t$  the standard Schwarzschild time.

Concretely, the Vaidya metric takes the form
\begin{equation}
    {\rm d}s^2=-\left(1-\frac{2 M(w)}{r}\right){\rm d}w^2+ 2 s\,{\rm d}w \, {\rm d}r+r^2{\rm d}\Omega_{S^2}^2\,,
\label{eq:Vaidya}
\end{equation}

with $s$ being the sign of the
derivative of the mass function $M(w)$, $s\equiv \text{sign}[M'(w)] $, and with ${\rm d}\Omega_{S^2}^2={\rm d}\theta^2+\sin^2\theta \, {\rm d}\phi^2$. In the following, we will assume  $M(w)$ to be a monotonic function of $w$, in such a way that $M'(w)$ has definite sign.

The metric \eqref{eq:Vaidya} describes an absorbing black hole by taking  $M'(w)> 0$ with $w = t+r_*$ (ingoing Vaidya), while it describes an emitting black hole if  $M'(w)< 0$ with $w = t-r_*$ (outgoing Vaidya).

The Vaidya BH metric \eqref{eq:Vaidya} solves the non-vacuum  Einstein equations
\begin{equation}
    R_{\mu\nu}-\frac{1}{2}R\,g_{\mu\nu}= 8\pi T_{\mu\nu }\,,
\label{eq:Einstein}
\end{equation}
where $R_{\mu\nu}$ and $R$ are the Ricci tensor and Ricci scalar, respectively, while $T_{\mu\nu}$ is the stress-energy tensor (SET) 
\begin{equation}
    T_{\mu\nu}=\frac{| M'(w)|}{ 4 \pi r^2}\partial_\mu w  \partial_\nu w\,.
\label{Vaidya_SET}
\end{equation}
The latter describes a pure radiation field (e.g., photons or gravitons) in the geometrical optics limit. 

The Vaidya BH presents a nontrivial causal structure, related to its departure from stationarity \cite{Nielsen_2010, 2014Galax...2...62N, RelativistToolkit}. Unlike in the static case, the apparent horizon (AH), defined as the hypersurface of vanishing expansion, i.e.~the boundary at which a congruence of null geodesics starts focusing into a trapped region, does not coincide with the event horizon (EH).
For a null congruence of geodesic curves, the outgoing null expansion $\Theta_{\text{outgoing}}$ coincides with the fractional variation of the cross-sectional area $\Sigma$ of the null congruence itself, with respect to an affine parameter $l$: 
\begin{equation}
    \Theta_{\text{outgoing}} =\frac{1}{\Sigma}\frac{\delta \Sigma}{\delta l}\,.
\end{equation}
In our case, this simply reads \cite{Nielsen_2010}:
\begin{equation}
    \Theta_{\text{outgoing}} = \frac{1}{r}\left(1-\frac{2 M(w)}{r}\right)\,.
\end{equation}
The condition $\Theta_{\text{outgoing}}=0$ therefore provides the location of the apparent horizon, namely $r_{\text{AH}}=2 M(w)$. 

The global event horizon, instead, cannot be defined locally. However, for a constant rate of increasing/decreasing mass, i.e.~$M(w)= M_0+M'\,(w-w_0)$, with $M'=\text{const}$, its location can be computed quite easily. 

To do so, it is convenient to move to a different coordinate frame, in which  the causal structure becomes completely manifest and, at the same time,  the equations for the perturbations take  a  simpler form, as we will discuss in the next section.

Consider first the transformation
\begin{equation}
\begin{cases}
    w&\rightarrow \quad W\equiv \int \frac{{\rm d}w}{2M(w)}\,, \\
    r&\rightarrow  \quad x\equiv \frac{r}{2M(w)}\,,
\end{cases}
\label{coordinate_transformation}
\end{equation}
which rescales both the radial coordinate and the null time. We can then trade the rescaled null time with a timelike coordinate defined as $T= W-s \, x_*$,\footnote{The time coordinate should not be confused with the trace of the SET, which we will denote below, when needed, with ${T^\lambda}_\lambda$.} where
\begin{equation}
x_*=\int  \frac{{\rm d}x}{f(x)} 
\label{eq:tortoisex}
\end{equation}
 is a generalized tortoise coordinate with
\begin{equation}
    f(x)=1-\frac{1}{x}-4|M'| x\,.
\end{equation}
In these coordinates, the Vaidya metric $g_{\mu\nu}$, defined in \eqref{eq:Vaidya},  
becomes conformal to a static metric $\Tilde{g}_{\mu\nu}$, i.e.,
\begin{equation}
g_{\mu\nu}=4 M(w)^2\Tilde{g}_{\mu\nu} ,
\label{eq:tildeg}
\end{equation}
and the line element reads
\begin{equation}
    {\rm d}s^2=4 M(w)^2\left[-f(x){\rm d}T^2+\frac{1}{f(x)}{\rm d}x^2+x^2{\rm d}\Omega_{S^2}^2 \right].
\label{conformal_metric}
\end{equation}
The location of the EH is now evident in this diagonal form of the metric from the condition $g^{xx}=0$, which corresponds, in physical coordinates, to 
\begin{equation}
\begin{split}
    r_{\text{EH}} & = 2M(w)\left(\frac{1-\sqrt{1- 16|M'|}}{8| M'|}\right)\\
    &=2M(w)\left(1+ 4 |M'|+\mathcal{O}(M'^2)\right)\,,
    \label{Event_Horizon}
\end{split}
\end{equation}
where in the second line we have expanded for small values of the mass derivative $M'$ and kept terms only up to linear order. Note that the EH always lies outside the AH. 

Notice also that the condition $g^{xx}=0$ yields at the same time an external ``cosmological horizon'' (CH), far away from $r_{\text{EH}}$ and located at  
\begin{equation}
r_{\text{CH}}=2M(w)\left(\frac{1+\sqrt{1-16 |M'|}}{8 |M'|}\right)\,.
\label{Cosmological_Horizon}
\end{equation}
This is formally similar to the cosmological horizon present in  Schwarzschild-de Sitter spacetime, with the mass derivative playing the role of a positive cosmological constant.

The choice of considering a constant rate of change of the BH mass, which we will adopt in the following, is physically relevant e.g. 
for accreting BHs (in which case the mass derivative is approximately given by the Eddington rate on timescales much longer than the ringdown).
For BHs whose mass changes  due to their own ringdown, the mass derivative can also be considered approximately constant on scales shorter than the ringdown decay time, and  order corrections depending on the higher time derivatives of the mass can in principle also be included perturbatively. 

Finally, note that the condition for the simultaneous existence of the two horizons of Eqs.~(\ref{Event_Horizon}) and (\ref{Cosmological_Horizon}) is $|M'|<1/16$.
We will restrict to this case hereafter.
 This choice is again motivated by the physical assumption that the mass evolution is a subleading effect in the perturbation dynamics.

\section{Perturbation equations}
\label{pert_eqs_Vaidya}

In this section, we derive the equations of motion for massless scalar, electromagnetic and (axial) gravitational perturbations of the Vaidya spacetime.
Thanks to the spherical symmetry of the background geometry \eqref{eq:Vaidya}, it will be convenient to adopt spherical coordinates and decompose the field perturbations in spherical harmonics.
We will work under the simplifying assumption that the mass rate $M'$ is constant. This will allow us to relate the $M' = \text{const}$ Vaidya metric   to a static one through a conformal transformation. As a byproduct,  the  equations for the perturbations will be time independent in the new coordinates, as we will show explicitly. This fact will  allow  us to study the solutions in the frequency domain, mirroring the  familiar case of perturbations around stationary BHs.

\subsection{Generalities and scalar case}
\label{scalar_equation}
As an illustrative example, let us start by considering the case of a test massless scalar field.
The dynamics  is captured  by the Klein--Gordon equation
\begin{equation}
    \Box \Phi = 0\,,
\end{equation}
where we have defined the d'Alembert operator $\Box =g^{\mu\nu}\nabla_\mu\nabla_\nu$, indicating with $\nabla$ the covariant derivatives.
 
Under a general conformal transformation of the metric, $g_{\mu\nu} = A^2 \Tilde{g}_{\mu\nu}$, a generic field $\varphi$ transforms as
\begin{equation}
    \varphi =A ^\chi\Tilde{\varphi}\,,
\end{equation}
with $\chi$ being the conformal weight. 
Choosing $\chi = -1$ for the scalar field $\Phi$, one obtains \cite{Wald:1984rg}
\begin{equation}
    \Box \Phi =\left(\Tilde{\Box} -\frac{\Tilde{R}}{6}\right)\,\Tilde{\Phi}=0 , 
    \label{KG_conformal}
\end{equation}
For the Vaidya metric \eqref{eq:Vaidya}, the Ricci scalar $R$ is proportional to the trace of the SET, which is a pure radiation field,
and it therefore vanishes, $R = 0$. However, it is nonzero for the conformal metric   $\Tilde{g}$ in \eqref{eq:tildeg}, which corresponds to choosing $A=2M(w)$ and yields $\Tilde{R}= 24 |M'|/x$.  Introducing the decomposition 
\begin{equation}
\Tilde{\Phi}(T,x,\theta,\phi) = \sum_{\ell m} \int\frac{{\rm d}\Omega}{2\pi} {\rm e}^{-i \Omega T} \frac{u(x)}{x}
Y_{\ell m}(\theta,\phi)\,, 
\end{equation}
the scalar equation reduces to
\begin{equation}
    \left[\frac{{\rm d}^2}{{\rm d}x_*^2}+\Omega^2-  f(x) \left(\frac{\ell(\ell +1)}{x^2}+\frac{1}{x^3}\right)\right]u(x)=0\,,
    \label{KG_conformal_2}
\end{equation}
where we used the tortoise coordinate $x_*$ defined in Eq.~\eqref{eq:tortoisex}. Note that, for ease of  notation, we dropped from $u(x)$  the dependence on the spherical harmonic quantum number $\ell$.

\subsection{Electromagnetic perturbations}
\label{em_equations}

Electromagnetic  perturbations are  described by the Maxwell equations
\begin{equation}
g^{\alpha\mu}\nabla_\alpha F_{\mu\nu}=0\,, 
    \label{Maxwell_eq}
\end{equation}
where $F_{\mu\nu}$ is the field strength, satisfying $\nabla_{[\alpha}F_{\mu\nu]}=0$. 
Unlike the Klein--Gordon equation, the Maxwell equations \eqref{Maxwell_eq} are invariant under conformal transformations, thanks to the scale invariance of electromagnetism in four spacetime dimensions (as it can be explicitly verified by choosing conformal weight $\chi = 0$). Therefore, we can directly consider the equations on the stationary geometry $\Tilde{g}_{\mu\nu}$, with the four-potential $A_\mu$ transforming trivially as $A_\mu=\Tilde{A}_\mu$.
Separating the field in axial (odd) and polar (even) parts according to their transformation rules under parity, which acts in polar coordinates  as  
$\theta\rightarrow \pi-\theta$, $\phi\rightarrow \phi+\pi$, we shall write
\begin{equation}
\begin{split}
        A_\mu \,{\rm d}x^\mu& =  \left(A_\mu^{\text{(polar)}}+A_\mu^{\text{(axial)}}\right){\rm d}x^\mu \\
        & = {\rm e}^{-i\Omega T}\Big[h(x)\,Y_{\ell m}\,{\rm d}T+e(x) \,Y_{\ell m}\,{\rm d}x \\
        &\qquad +\left(a(x) {\varepsilon_i}^j\,\partial_jY_{\ell m}+k(x)\,\partial_iY_{\ell m}\right){\rm d}x^i\Big]\,,
\label{four_potential}
\end{split}
\end{equation}
with $\varepsilon$ being the Levi-Civita symbol on the two-dimensional Euclidian sphere. In addition, we will make the gauge choice  $k(x)=0$. 

The Maxwell equation $\nabla^\mu F_{\mu \phi}=0$ immediately yields the equation of motion of the  odd electromagnetic degree of freedom:
\begin{equation}
    \left[ \frac{{\rm d}^2}{{\rm d}x_*^2}+\Omega^2-  f(x) \frac{\ell(\ell +1)}{x^2}\right] a(x)=0\,.
    \label{EM_oddP}
\end{equation}
On the other hand, from  $\nabla^\mu F_{\mu \theta}=0$, we obtain
\begin{equation}
    h(x) = \frac{i}{\Omega}f(x)\frac{{\rm d}}{{\rm d}x}\left(f(x) \,e(x)\right)\,.
\end{equation}
Substituting this into $\nabla^{\mu} F_{\mu x}=0$, 

and introducing the variable $q(x) = f(x)e(x)$, one gets the equation for the electromagnetic polar degree of freedom:
\begin{equation}
    \left[\frac{{\rm d}^2}{{\rm d}x_*^2}+\Omega^2-  f(x) \frac{\ell(\ell +1)}{x^2}\right]q(x)=0\,.
    \label{EM_evenP}
\end{equation}
Note that the equation is identical to the one in Eq.~\eqref{EM_oddP} for the axial mode. This means in particular that the two sectors are isospectral, i.e.~they share the same set of quasi-normal modes. This result is in fact more general than the case of constant $M'$, which we focused on here. We will explicitly verify this in Appendix~\ref{perturbation_eqs} for arbitrary choices of $M(w)$, where we also discuss the connection with electric-magnetic duality.

\subsection{Gravitational perturbations}
\label{sec:gravpts}

Analogously to the electromagnetic case,  we perform a similar split in axial and polar components for gravitational perturbations. This will ensure that the  dynamics of the two sectors is decoupled at the level of the linearized Einstein equations.
Note that, thanks to the  structure of the SET of Eq.~(\ref{Vaidya_SET}), the perturbations of the radiation field couple only to the polar gravitational perturbations, and do not affect the dynamics of the axial gravitational sector. In fact, since $T_{\mu\nu}\propto \partial_\mu w \partial_\nu w$, its fluctuations $\delta T_{\mu\nu}$ can only transform evenly under parity. Hence, for simplicity, we will focus below on the axial gravitational sector only, and leave the analysis of the polar perturbations for a future study.

To obtain  the master equation for the axial modes, one can proceed similarly to the derivation of the Regge--Wheeler equation for odd perturbations of Schwarzschild BHs. In the following,  we will work in the ``Einstein frame'' \cite{faraoni1998conformal}, i.e.~we will consider the equations of motion for the conformal metric $\Tilde{g}_{\mu\nu }$. We will show in Appendix~\ref{equation derivation} that this approach is completely equivalent to the derivation of the Regge--Wheeler equation in the Jordan frame, i.e.~in terms of the Vaidya metric $g_{\mu\nu }$, satisfying the usual Einstein equations.  In the following expressions, we will use the notation $s = \text{sign}(M')$,   $M'$ and $|M'|$ to stress the  distinction between   situations where the mass derivative appears with its sign or in absolute value.

The equations of motion for the conformal metric $\Tilde{g}$ read \cite{Wald:1984rg}
\begin{multline}
\Tilde{R}_{\mu\nu}-2 \nabla_\mu\nabla_\nu \ln A-\Tilde{g}_{\mu\nu}\,\Box \ln A \\
     +2 \nabla_\mu \ln A \nabla_\nu \ln A-2 \Tilde{g}_{\mu\nu}\nabla_\alpha \ln A \,\nabla^\alpha \ln A \\
     =8\pi T_{\mu\nu}\,,
\end{multline}
where $A=2M(w)$ and 
where the SET is 
\begin{equation}
    T_{\mu\nu}{\rm d}x^\mu{\rm d}x^\nu= \frac{|M'|}{4 \pi x^2}\left[{\rm d}T^2+\frac{2s}{f(x)}{\rm d}T {\rm d}x+
    \frac{{\rm d}x^2}{f(x)^2}\right],
\end{equation}
 with the logarithmic gradient of the conformal factor reducing to
\begin{equation}
    \nabla_\mu \ln A \,{\rm d}x^\mu= 2 |M'| \left({\rm d}T+\frac{s}{f(x)}{\rm d}x\right)\,.
\end{equation}

Let us consider now a perturbation of the metric in the Einstein frame, i.e.~let us write $\Tilde{g}+\Tilde{h}$, and split the perturbation in  axial and  polar sectors as $\Tilde{h}=\Tilde{h}^{\text{(ax)}}+\Tilde{h}^{\text{(pol)}}$. 
Since $\Tilde{h}^{\text{(ax)}}$ and $\Tilde{h}^{\text{(pol)}}$ are decoupled at linear order,  it is consistent to set $\Tilde{h}^{\text{(pol)}}$ to zero and  focus only on $\Tilde{h}^{\text{(ax)}}$.

In the $(T,x,\theta,\phi)$ coordinates and in the Regge--Wheeler gauge \cite{PhysRev.108.1063}, we shall write
\begin{equation}
\begin{split}
\Tilde{h}^{(\text{ax})}_{\mu\nu}&=
\sum_{\ell m} \int\frac{{\rm d}\Omega}{2\pi}
\left(
\begin{array}{cccc}
 0 & 0 & -\frac{h_0(x)}{\sin \theta}\, \partial_\phi  & \sin \theta \,h_0(x) \, \partial_\theta  \\
 0 & 0 &  -\frac{h_1(x)}{\sin \theta}\,\partial_\phi  & \sin \theta \,h_1(x) \, \partial_\theta \\
\text{Sym} & \text{Sym} & 0 & 0 \\
\text{Sym} & \text{Sym} & 0 & 0 \\
\end{array}
\right) \\
& \quad \times  {\rm e}^{-i\Omega T} Y_{\ell m}(\theta ,\phi ) \,.
\end{split}
\label{AxialPerturbation}
\end{equation}
The three nontrivial components of the linearized Einstein equations are: 
\begin{multline}
\frac{f(x)}{x}\left[h_0''(x)+i\Omega \,h_1'(x)\right]
    +\frac{2i\Omega}{x}\left[f(x)+2|M'|x\right]h_1(x)\\
     + \frac{4 |M'|}{x}h_0'(x)-\left[\frac{\ell(\ell+1)}{x^2}-\frac{2}{x^3}\right]h_0(x)=0 \, ,
\end{multline}
\begin{multline}
h_0'(x)-\frac{2}{x}h_0(x)\\
    +s\left[  i \Omega -f(x)\frac{\ell(\ell+1)-2}{(4 |M'|-i\Omega )x^2} \right] h_1(x)=0 \, ,
\end{multline}
\begin{equation}
f(x)^2x^2 h_1'(x)+sf(x)h_1(x)+(i\Omega- 4 M')h_0(x)=0 \, .
\end{equation}

\noindent From the last equation, we can isolate $h_0(x)$ and substitute it in the second one. Then, after introducing  the master variable $Q(x)$, defined  as $h_1(x)=F(x)Q(x)$, with
\begin{equation}
    F(x)=-\frac{x^{\frac{5}{4}}}{f(x)^{\frac{3}{4}}}\,\exp{\left(\frac{\arctan\left(\frac{8|M'|-1}{\sqrt{16|M'|-1}}\right)}{2\sqrt{16|M'|-1}}\right)}\,,
\end{equation}
we find the following    equation for $Q(x)$:
\begin{equation}
    \left[\frac{{\rm d}^2}{{\rm d}x_*^2}+\Tilde{\Omega}^2- f(x) \left(\frac{\ell(\ell +1)}{x^2}-\frac{3}{x^3}\right)\right]Q(x)=0\,,
\end{equation}
with the shifted frequency $\Tilde{\Omega}=\Omega+2i M'$ (note that here the mass derivative must be taken with its positive or negative sign).
 \\

\subsection{Master equation}
\begin{figure}[t]
          \centering
\includegraphics[width =0.5\textwidth]{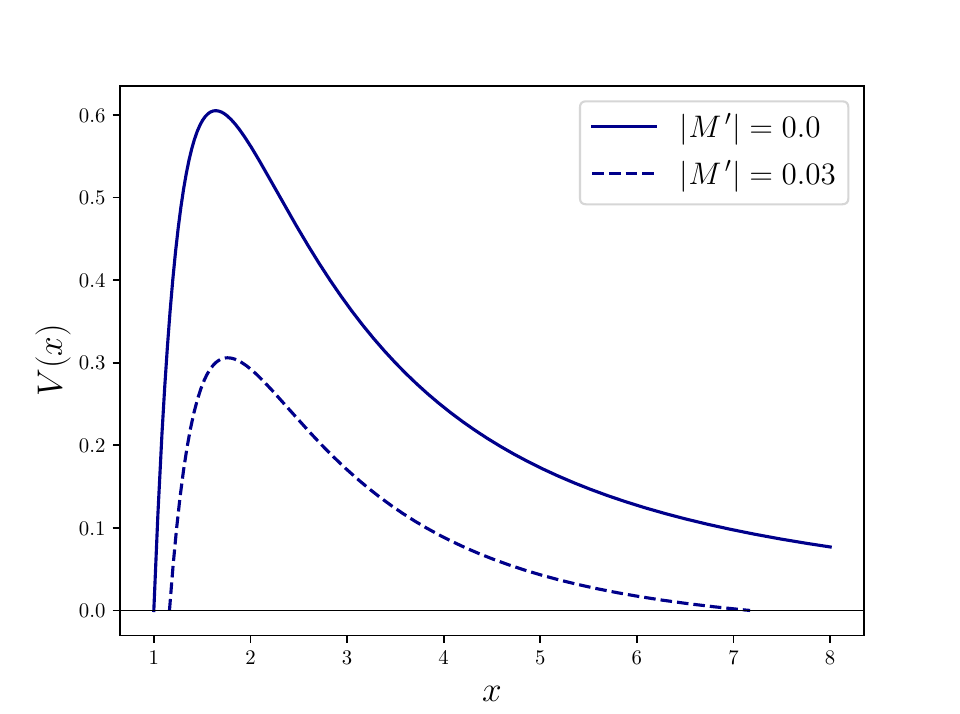}
\caption{Gravitational potential ($\sigma=-3$) for $\ell=2$. Note that a non-vanishing mass derivative $M'$ shifts the event horizon outwards, while causing the appearance of an external ``cosmological horizon''.}
    \label{fig:V_RW}
\end{figure}

In summary, one can write the equations for all kinds of perturbations (scalars, vectors and axial tensors) as a single master equation for a suitable master variable $R(x)$:
\begin{equation}
    \left[\frac{{\rm d}^2}{{\rm d}x_*^2}+\Tilde{\Omega}^2- \, f(x) \left(\frac{\ell(\ell +1)}{x^2}+\frac{\sigma}{x^3}\right)\right]R(x)=0\,,
    \label{RW_conformal}
\end{equation}
where $\Tilde{\Omega}=\Omega$ in the scalar and electromagnetic case, and $\Tilde{\Omega} = \Omega+2iM'$ in the axial gravitational  case. The spin parameter $\sigma$ reads respectively $1,0,-3$ for scalar, electromagnetic and axial gravitational  perturbations.
Unlike in the static case ($M'=0$), the potential  
\begin{equation}
V(x) = f(x) \left(\frac{\ell(\ell +1)}{x^2}+\frac{\sigma}{x^3}\right)
\label{eq:VxV}
\end{equation}
vanishes in two points, namely at the EH and at the CH, see Fig.~\ref{fig:V_RW} for an example.

Note that the main point of our approach is the time-independent form of Eq.~\eqref{RW_conformal}, which can be solved as a boundary value problem in the frequency domain, as we will discuss in the following. 
In \cite{Shao:2004ws, Abdalla:2006vb,Chirenti:2010iu, Chirenti:2011rc, Lin:2021fth, Redondo-Yuste:2023ipg}, time-domain methods have been instead used to solve the wave equation with time-dependent potential (see Appendix \ref{perturbation_eqs} for the expression of the perturbation equations in Eddington--Finkelstein coordinates), given  initial conditions and a functional form for the mass evolution function $M(w)$.
%
In most of those works, the time-evolution is performed in double null coordinates $u,v,\theta,\phi$, where $u$ and $v$ are the retarded and advanced time that we already defined, starting from initial conditions given by a Gaussian wave packet.
This procedure allows for a generic time evolution for the mass, but the extraction
of the QNM frequencies from the time domain signal can only be performed for the dominant mode, and is generally less accurate than in our frequency domain approach.
In the next section, we present the computation of the QNMs in the frequency domain from Eq.~\eqref{RW_conformal}, showing that the full spectrum of frequencies can be obtained in this framework.

\section{Quasi-Normal Modes}
\label{QNMs}

\subsection{Eikonal approximation}
\label{Eikonal_Corresp}

We start here by deriving the Vaidya QNM frequencies in the geometrical optics limit---also known as eikonal limit---while a full numerical computation of the spectrum by means of the continued-fraction method will be presented later on in Sec.~\ref{sec:freqdom}. 
Besides providing a first  approximation for the spectrum, the eikonal limit can shed light on the correspondence between the frequencies and the parameters of the unstable photon geodesics at the light-ring. For stationary BHs in GR, it was found \cite{Cardoso_2009, Solanki_2022, Konoplya_2017} that the real part of the eikonal QNMs corresponds to multiples of the orbital frequency, while the imaginary part is related to the Lyapunov exponent, which characterizes the instability timescale of the orbit. 
It is interesting to check whether there are examples of non-stationary BHs for which such correspondence can be recovered.

The potential in  Eq.~\eqref{eq:VxV}, in the eikonal limit $\ell\gg1$, reads
\begin{equation}
    V(x)=\Big(1-\frac{1}{x}-4|M'|\,x\Big)\frac{\ell^2}{x^2}+\mathcal{O}(\ell^0)\, .
\label{eikonal_potential}
\end{equation}
It has two stationary points, a global maximum and a global minimum. The latter is located outside the cosmological horizon. The former corresponds to the light-ring, which is  shifted with respect to the Schwarzschild one. Let us consider a small linear evolution of the mass function, namely $M' \rightarrow \epsilon M'$, where $\epsilon$ is a (positive) book-keeping parameter, which we will take to be small enough such that   $\epsilon M'w\ll M_0$. The location of the light-ring $r_{lr}$, at linear order in $\epsilon$, evolves in time as 
\begin{equation}
\begin{split}
r_{lr}&=3M(w)\left[1+3|M'|\epsilon +\mathcal{O}(\epsilon^2)\right]=\\
        &=3M_0\left[1+3\left(1+s\frac{w-w_0}{M_0}\right)|M'|\epsilon 
+\mathcal{O}(\epsilon^2)\right]\,.
\end{split}
\end{equation}
This radius can also be obtained from the geodesics equation for photons \cite{Mishra:2019trb,Song:2021ziq,Chen_2022, Koga:2022dsu, Vertogradov:2023otc}. 
As already mentioned, 
the metric $\Tilde{g}_{\mu\nu }$ of Eq.~(\ref{conformal_metric}),

 which is conformally related to the Vaidya metric, is static and admits the stationarity Killing vector $\xi=\partial_T$. 

The equation for null geodesics is the same for both $g_{\mu\nu }$ and $\Tilde{g}_{\mu\nu }$, and reads
\begin{equation}
\begin{split}
0= \Tilde{g}_{\mu\nu}k^\mu k^\nu=&-\left(1-\frac{1}{x}- 4|M'| x\right)\frac{{\rm d}T}{{\rm d}l}+\\
    &+\frac{1}{1-\frac{1}{x}- 4|M'| x}\frac{{\rm d}x}{{\rm d}l}+x^2\frac{{\rm d}\phi}{{\rm d}l}\,,
\end{split}
\end{equation}
with $l$ being an affine parameter and $k^\mu$ a null vector. 

One can then use the conserved quantities associated with the Killing vector $\xi$ and the axial Killing vector $\kappa=\partial_\phi$,
\begin{equation}
\begin{split}
    E&=-\xi_T \, ,\\
    L&=\kappa_\phi\,,
\end{split}
\end{equation}
to obtain the equation
\begin{equation}
    \frac{{\rm d}x}{{\rm d}l}=E-V_{\text{eff}}(x)\,.
\end{equation}
The effective potential matches the one appearing in Eq.~(\ref{eikonal_potential}): 
\begin{equation}
    V_{\text{eff}}(x)=\Big(1-\frac{1}{x}-4| M'|\,x\Big)\frac{L^2}{x^2}\,.
\end{equation}
This fact shows that the eikonal correspondence between QNMs and null geodesics, which holds for stationary BHs in GR, is  recovered in the linear evolution limit of spherically symmetric dynamical BHs described by the Vaidya geometry. 

The quasi-normal frequencies can then be estimated analytically in the eikonal approximation.  In more detail, we have
\begin{equation}
\begin{split}
    &\Tilde{\Omega}^E_R= \sqrt{\frac{V(x_M)}{\ell^2}}\left(\ell+\frac{1}{2}\right)+\mathcal{O}(\ell^{-1}),\\
    &\Tilde{\Omega}^E_I=-\left.\frac{{\rm d}x}{{\rm d}x_*}\sqrt{\frac{V''(x)}{2V(x)}}\right|_{x_M}\left(n+\frac{1}{2}\right)+\mathcal{O}(\ell^{-1})\,,
\end{split}
\label{eikonal_frequencies}
\end{equation}
where $\Tilde{\Omega}^E_R$ and $\Tilde{\Omega}^E_R$ are the real and imaginary part of the QNM frequencies computed in the eikonal limit and $x_M$ is the position of the maximum of the potential.

Considering again a small mass derivative, we have, at leading order,
\begin{equation}
\begin{split}
     &\Tilde{\Omega}^E_R= \frac{2}{\sqrt{3}}\left(\ell+\frac{1}{2}\right)\left[\frac{1}{3}-3|M'|\,\epsilon+\mathcal{O}(\epsilon^2)\right] , \\
    &\Tilde{\Omega}^E_I=-\frac{2}{\sqrt{3}}\left(n+\frac{1}{2}\right)\left[\frac{1}{3}-4|M'| \,\epsilon+\mathcal{O}(\epsilon^2)\right]\,.
    \label{eikonal_frequencies_expanded}
\end{split}
\end{equation} 
These expressions can be directly related to the aforementioned parameters of the unstable circular photon orbit \cite{Cardoso_2009, Solanki_2022, Konoplya_2017}. 

\subsection{Numerical analysis}
\label{sec:freqdom}

\begin{figure*}[t]
          \centering
\includegraphics[width = 0.8\textwidth]{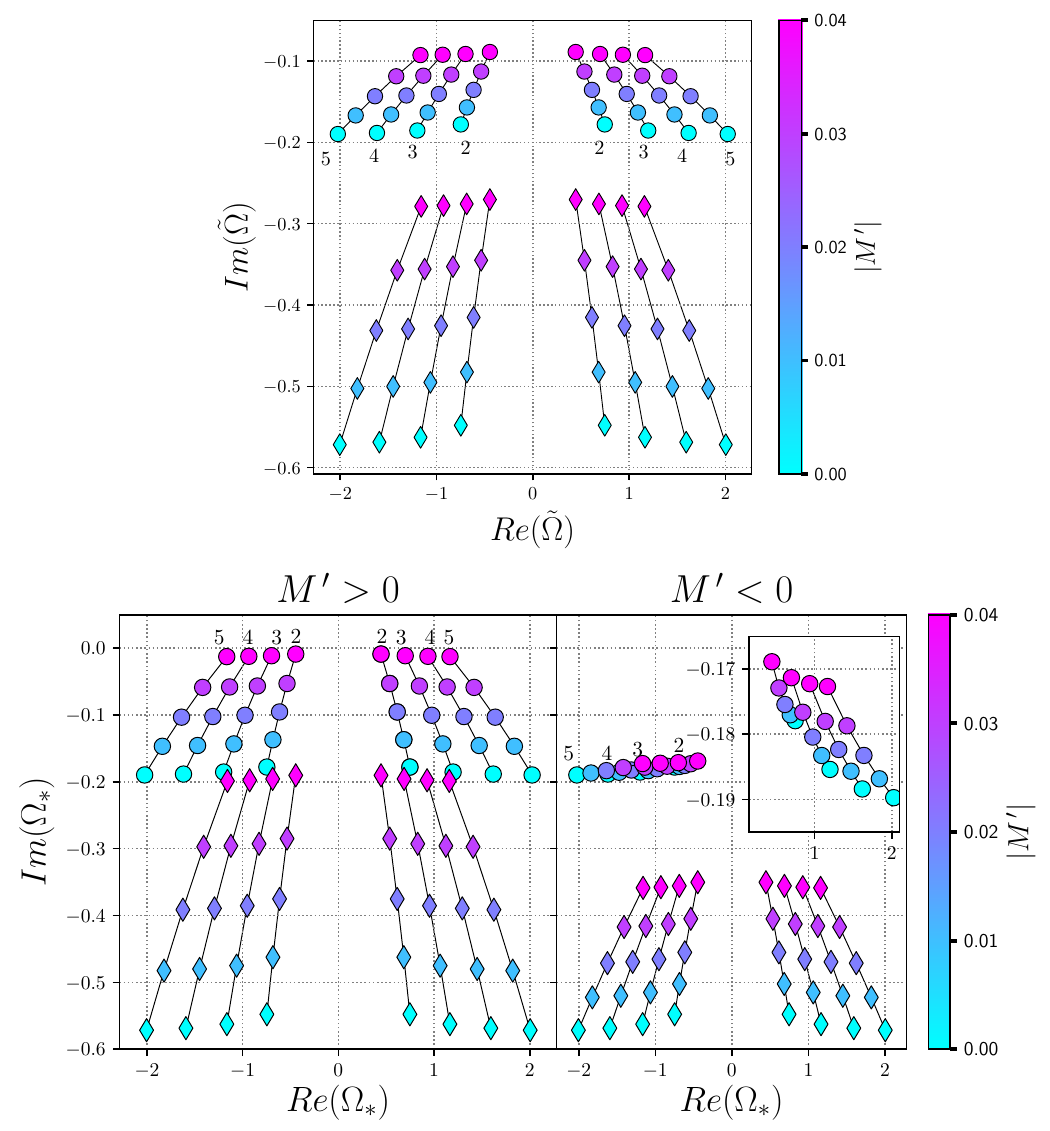}
 \caption{QNM rescaled frequencies in the complex plane. The fundamental modes are presented with round points, while the first overtones with diamond-shaped symbols. The small black numbers inside the plot indicate the angular momentum number $\ell$, which we vary within the interval $\ell\in [2,5]$. The color code indicates the modulus of the mass derivative. The upper panel represents the frequencies $\Tilde{\Omega}$ in the master equation Eq.~(\ref{RW_conformal}): at this level there is no difference between the increasing and decreasing mass cases. The lower panels represent instead the frequencies $\Bar{\Omega}$ which are shifted according with the conformal weight of the gravitational perturbation field: here the spectrum is instead different in the cases in which the mass is increasing (left panel) or decreasing (right panel). Note that for large $M'$ in the lower-left panel the frequencies approach the unstable regime.         
          }
\label{fig:ComplexPlane}
\end{figure*}
In order to find the full frequency spectrum, we solve  Eq.~(\ref{RW_conformal}) using Leaver's continued-fraction method. 
The procedure that we adopt for the Vaidya BH is similar to the Schwarzschild-de Sitter case, in which a study of QNMs with the same approach has been carried out in e.g.~\cite{PhysRevD.106.124004, Yoshida_2004, Zhidenko:2003wq}. 
This method follows from the more general Frobenius procedure for second-order linear differential equations, which allows for finding solutions in terms of infinite power series. 

Two of the four singular points of the wave equation (\ref{RW_conformal}) (the other two being $x =0,\infty$) correspond to the positive zeros of the function $f(x)$, the first being the EH $x_H$, and the second being the CH   $x_C\sim 1/|M'|$.
The QNMs can be defined as solutions to the wave equation that are purely ingoing at the EH and purely outgoing at the CH.
We can impose these boundary conditions in terms of the tortoise coordinate 
\begin{equation}
\begin{split}
    x_* & =\int\frac{{\rm d}x}{f(x)}  \\
    & =\frac{1}{4 |M'| (x_C-x_H)}\Big[x_H\ln(x-x_H)   -x_C\ln(x-x_C)\Big].
\end{split}
\end{equation}
They read 
\begin{equation}
    \begin{split}
        &R(x_*)\xrightarrow{x \rightarrow x_C}{\rm e}^{i\Tilde{\Omega} x_*} , \\
        &R(x_*)\xrightarrow{x \rightarrow x_H}{\rm e}^{-i\Tilde{\Omega} x_*}\,.
    \end{split}
\label{eq:Rbcs}
\end{equation}
Note that we can safely impose these outgoing/ingoing conditions at the horizons in terms of the frequency $\Tilde{\Omega}$, as the latter is related to $\Omega$ by a shift in the imaginary part, which only affects the amplitude of the mode.

\begin{figure}[t]
 \resizebox{\hsize}{!}{
\includegraphics[width =0.1\textwidth]{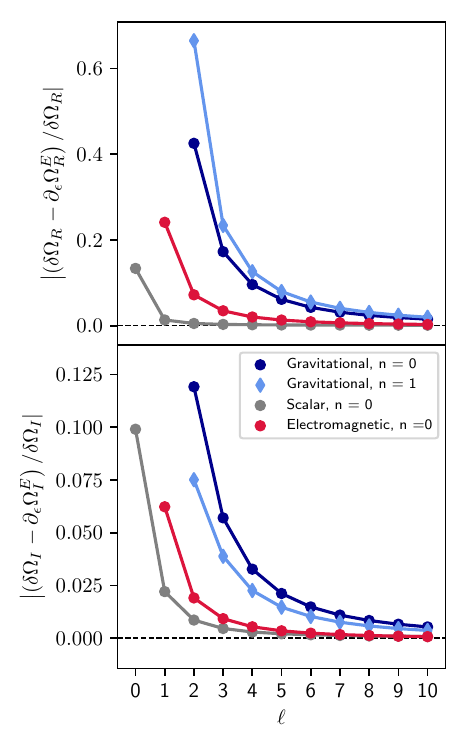}
          }
\caption{Relative error between the linear coefficients $\delta\Omega$ computed with the Leaver method and those computed analytically in the eikonal limit. It can be observed that, as expected, the eikonal approximation becomes better as $\ell$ increases.}
          \label{fig:LinCoefficients}
\end{figure}

The solution to Eq.~\eqref{RW_conformal}, subject to the boundary conditions~\eqref{eq:Rbcs} can then be expressed as the product of a diverging function at the horizons, and a convergent infinite power series in the interval $(x_H,x_C)$. We can therefore use the ansatz 
\begin{equation}
    R(x)=(x-x_C)^{\rho\,\eta\,x_C} (x-x_H)^{\rho\,\eta\,x_H}\,S(x)\,,
\end{equation}
where we defined $\rho = -i\Omega $ and $\eta = \frac{1}{4|M'|(x_C-x_H)}$, and where $S(x)$ is the Frobenius series
\begin{equation}
    S(x)= \sum_{n=0}^{\infty} a_n\left(\frac{x-x_H}{x}\right)^n\,.
\end{equation}
Inserting this ansatz into the master equation, one obtains the following five-term recurrence relation among the coefficients $a_n$ 
\begin{equation}
    \alpha_n \,a_{n+1}+\beta_n\,a_n+\gamma_n\,a_{n-1}+\delta_n\,a_{n-2}+\zeta_n\,a_{n-3}=0\,.
\end{equation}
The explicit expressions for these coefficients, which are quite lengthy, are presented in Appendix \ref{5t relation}. 

This five-term relation can be reduced numerically through Gaussian elimination in two steps. One first define the new coefficients as
\begin{equation}
\begin{split}
    &\alpha'_n=\alpha_n \,,\\
    &\beta'_n = \beta_n-\frac{\alpha'_{n-1}}{\delta'_{n-1}}\,\zeta_n\,, \\
    &\gamma'_n = \gamma_n-\frac{\beta'_{n-1}}{\delta'_{n-1}}\,\zeta_n\,, \\
    &\delta'_n = \delta_n-\frac{\gamma'_{n-1}}{\delta'_{n-1}}\,\zeta_n\,.
\end{split}
\end{equation}
The same procedure can be repeated to obtain the three-term relation
\begin{equation}
    \alpha''_n \,a_{n+1}+\beta''_n\,a_n+\gamma''_n\,a_{n-1}=0\,.
\end{equation}
At this point, one can define the $n$-th ladder operator from the ($n+1$)-th one, 
 as
\begin{equation}
    A_n = \frac{\gamma''_n}{\beta''_n-\alpha''_n\,A_{n+1}}\,.
\end{equation}
Finally, the spectrum is obtained by finding the zeros of the Leaver function
\begin{equation}
    \mathcal{F}_{\text{Leaver}}(\Tilde{\Omega},\epsilon)=A_1-\frac{\beta''_0}{\alpha''_0}=0\,.
\end{equation}

The frequencies obtained in this way from the spectrum of the master variable in \eqref{RW_conformal} do not represent the physical spectrum. 
To obtain the physical spectrum, two effects  must be taken into account. First, the physical perturbations are related to the master variable by a factor $(2 M(w))^\chi$, where $\chi$ is the conformal weight of the fields. This effect  introduces a shift in the frequencies $\Tilde{\Omega}$. Second, because we performed a  Fourier transform in the dimensionless time variable $T$, the frequencies must be rescaled by an overall time-dependent factor. This second effect will be discussed in more detail in the next subsection. 

In the rescaled coordinates, and ignoring all factors depending on $x$ (which affect the amplitude and not the frequency), the time-dependent mass reads 
\begin{equation}
    M(w)^\chi\sim {\rm e}^{2\chi M' T}\,.
\end{equation}
The physical fields will then oscillate with frequency 
\begin{equation}
    \Omega_*\equiv\Omega+2 i \chi M'\,,
    \label{OmegaStar}
\end{equation}
where $\Omega = \Tilde{\Omega}$ for scalar and electromagnetic perturbations and $\Omega = \Tilde{\Omega}-2i M'$ for gravitational axial perturbations. Again, this is a purely imaginary shift, so it does not affect the boundary conditions of Eq.~(\ref{eq:Rbcs}).
The rescaled frequencies of the axial metric perturbations ($\chi = 2$), which represent the physical gravitational perturbations (in the axial sector), then oscillate  with frequency $\Tilde{\Omega}+2 i M'$. The overall shift of the gravitational spectrum in the complex plane is represented in Fig \ref{fig:ComplexPlane}. 
The fundamental modes are represented with round points, while the first overtones are indicated with diamond-shaped points. The color-code represents the absolute value of the mass-derivative. The upper panel shows the frequency spectrum $\Tilde{\Omega}$, i.e.~before the shift $2i\chi M'$ is introduced. This spectrum does not depend on the sign of the mass derivative. The lower panels show instead the spectrum $\Omega_*$, which differs when the mass increases or decreases.

\subsection{Small rate limit and unstable regimes}
For small values of the mass derivative the quasi-normal frequencies can be approximated as  
\begin{equation}
    \Omega_{n \ell}\simeq \Omega_{n \ell}^{(0)}+\delta\Omega_{n \ell}\,\epsilon\,,
\end{equation}
with $\Omega_{n \ell}^{(0)}$ being the Schwarzschild QNMs and $\delta\Omega_{n\ell}$ being a constant complex coefficients. 
This can be viewed as a Taylor expansion of the frequencies around $\epsilon = 0$, namely
\begin{equation}
    \Omega(\epsilon)=\Omega(0)+\left.\frac{\partial\Omega}{\partial\epsilon}\right|_{\epsilon=0}\epsilon+\mathcal{O}(\epsilon^2)\,.
\end{equation}
The first order correction can be obtained from a linear fit of the numerical results, and can also be estimated analytically in the eikonal limit from Eq.~(\ref{eikonal_frequencies}). 

One can   see from Fig.~\ref{fig:LinCoefficients} that the agreement between the numerical $\delta\Omega$ and the one computed with Eq.~(\ref{eikonal_frequencies}) gets better and better as $\ell$ increases, exactly as expected.

On the other hand, for high enough values of $|M'|$, the imaginary part of the quasi-normal frequencies can change sign due to the shift $2 i \chi M'$, leading to the onset of an instability. The sign of the conformal weight $\chi$ determines whether this unstable regime can be developed in the increasing or decreasing mass case.
Gravitational perturbations can grow unstable for $M'>0$. Indeed, it is evident from Fig.~\ref{fig:ComplexPlane} that the rescaled frequencies approach the limit of vanishing imaginary part for positive enough $M'$. On the other hand, scalar perturbations  transform with conformal weight $\chi = -1$, and therefore they can become unstable for negative enough values of $M'$. Electromagnetic perturbations are instead always stable. 

As previously stated, we are only interested in the range $|M'|<1/16$ for the mass derivative. However, values of $M'$ in this range can produce instabilities. In particular, the instability threshold for the fundamental $\ell=2$ gravitational mode is $M'_{\text{Tr}}\simeq 0.0420$, while for the fundamental $\ell = 0$ scalar mode  is $M'_{\text{Tr}}\simeq -0.0436$. 
The stability properties for different perturbations are summarized in Table~\ref{tab:instabilites}. 
\begin{table}[h!]
\centering
\resizebox{0.35\textwidth}{!}{
\begin{tabular}{|c|c|c|}
\hline

 & $M'>0$ & $M'<0$ \\
\hline
 Scalar & \cross & \tick \\
\hline
 Electromagnetic & \cross & \cross \\
\hline
 Gravitational axial & \tick & \cross \\
\hline
\end{tabular}
}
\caption{Possibility of developing an instability for different kind of perturbations in the two cases in which the BH mass is increasing and decreasing. The symbol \tick~signals that the quasi-normal frequencies can have an instability for a large enough $|M'|$. On the other hand, the symbol \cross~indicates that this is never the case.}
\label{tab:instabilites}
\end{table}

\subsection{Time-domain waveforms}

\label{time_domain}
 
In the following analysis we will just consider the small mass derivative case, in order to avoid possible instabilities. 
As already anticipated, in addition to including the shift in the imaginary part, in order to obtain the physical spectrum, we also have to account for the fact  that the physical frequencies  present an intrinsic time dependence determined by the evolution of the mass. 

The appearance of a time-depending factor in the physical frequencies can be seen from extracting the time evolution of the signal at $x\rightarrow x_C$. 
One has
\begin{figure}[t]
          \centering
\includegraphics[width =0.45\textwidth]{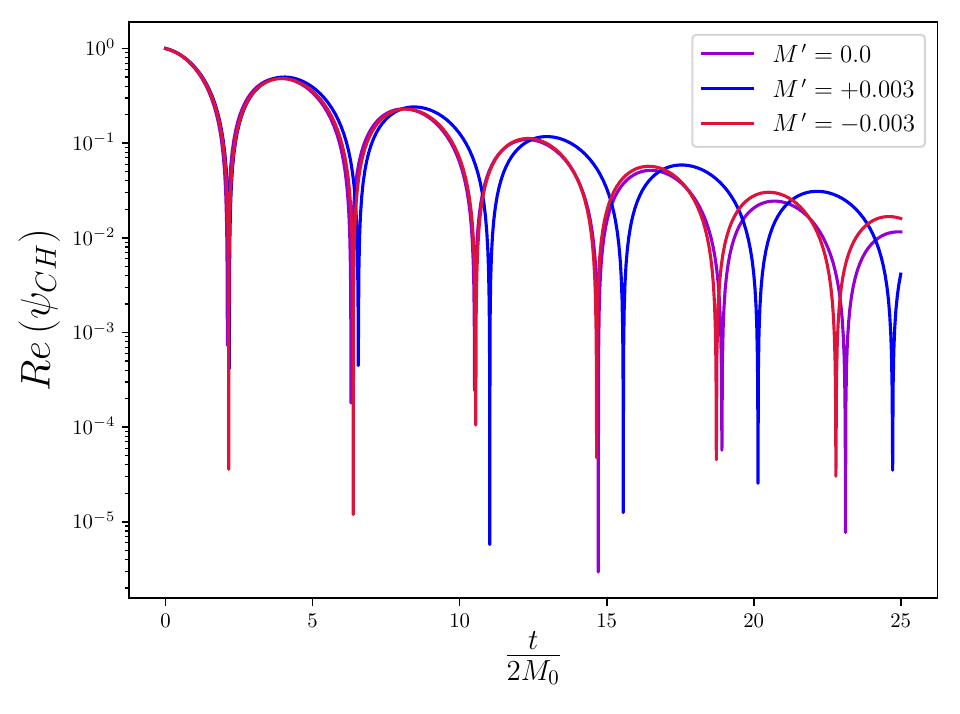}
 \caption{
Signal evolution at the cosmological horizon at leading order in $\epsilon$, for different mass derivative values}
\label{fig:Wf_evolution}
\end{figure}
\begin{equation}
\begin{split}
    \psi_{\rm CH}(T,x)&={\rm e}^{-i\Omega T}(x-x_C)^{-i\Omega \eta x_C}=\\
    &={\rm e}^{-i\Omega \int\frac{{\rm d}w}{2 M(w)}}{\rm e}^{-i\Omega x_*(x_C)}(x-x_C)^{-i\Omega \eta x_C}=\\
    &={\rm e}^{-i\Omega \int\frac{{\rm d}w}{2 M(w)}}(x-x_C)^{-i\Omega x_C (\eta-\eta)}=\\
    &={\rm e}^{-i\int \omega(w) {\rm d}w}\,,
\end{split}
\end{equation}
where in the last line we defined  the $w$-dependent frequency $\omega(w)$.
The same computation can be carried out close to the event horizon:
\begin{equation}
\begin{split}
    \psi_{\rm EH}(T,x)&={\rm e}^{-i\Omega T}(x-x_H)^{-i\Omega \eta x_H}=\\
     &={\rm e}^{-i\int \omega(w) {\rm d}w}(x-x_H)^{-2i\Omega \eta x_H}= \\
     &={\rm e}^{-i\int \omega(w) {\rm d}w}\,{\rm e}^{2 i\Omega\eta x_H\,\ln(2 M(w))}\times\\
&\qquad\times(r-r_H)^{-2i\Omega \eta x_H}= \\
     &={\rm e}^{-i(1-4\eta x_H M')\int \omega(w) {\rm d}w}(r-r_H)^{-2i\Omega \eta x_H}\,,
\end{split}
\end{equation}
where we used the relation
\begin{equation}
    \int\frac{{\rm d}w}{2 M(w)}=\frac{1}{2 M'}\ln(2 M(w))\,,
\end{equation}
which is valid in the $M'=\text{const}$ case.  
Note that if we introduce again the small parameter $\epsilon \ll 1$, the $w$-dependent frequency at the horizon gets an order $\epsilon$ overall correction with respect to the one measured at the outer horizon.
In fact, one has
\begin{equation}
    \eta \, x_H M'\epsilon = M'\epsilon + \mathcal{O}(\epsilon^2)\,.
\end{equation}

To summarize, two separate effects modify the QNM frequencies as a consequence of the BH mass evolution. The first is a `static' effect appearing already at the level of the rescaled coordinates, which  displaces the QNMs with respect to the Schwarzschild positions on the complex plane. The second is the time-dependent scaling of the physical frequency with the mass:
\begin{equation}
    \omega(w)=\frac{\Omega}{2 M(w)}\,.
    \label{physical_freq}
\end{equation}
At this point, one can  reconstruct the physical observable signal, which we show in Fig.~\ref{fig:Wf_evolution}.

Notice that the deviation from the static background signal is more evident in the increasing mass case, because the damping of the real part of the rescaled frequency in the complex plane adds up to the suppression of the physical frequency according to Eq.~(\ref{physical_freq}). On the other hand, the two effects are  competing in the decreasing mass case, canceling out at the beginning, before the time-dependent enhancement starts prevailing.

\section{Tidal response}
In the first part of the work, we have discussed the QNM spectrum of a Vaidya BH, assuming that the BH mass evolves at a constant rate. Here, we study a different type of effect, namely the tidal deformability of the BH induced by an external  perturbation.

In a relativistic context, the conservative  tidal response of a compact object is parametrized in terms of a set of coefficients, which are often referred to as Love numbers. It is well known that, as opposed to neutron stars or other  types of self-gravitating compact sources,  asymptotically-flat BHs in GR have vanishing  static Love numbers \cite{Fang:2005qq,Damour:2009vw,Binnington:2009bb,Gurlebeck:2015xpa}.  In the following, we wish to study how this result gets modified for a BH with varying mass, in the particular case of the Vaidya geometry \eqref{eq:Vaidya}.

\label{LoveNumbers}
\subsection{Love numbers}

Our starting point is the master equation \eqref{RW_conformal}. We will work under the assumption that $\vert M'\vert=\text{const}\ll 1$: in other words, we will treat  terms proportional to $M'$  as small corrections  to the Schwarzschild solution. To see how  the Vaidya corrections affect the unperturbed Schwarzschild result, it will be enough to keep terms only up to linear order in $M'$. We will thus  drop everywhere terms that are $\mathcal{O}(M'^2)$ or higher.

As opposed to the computation  of the QNMs  in Sec.~\ref{QNMs}, we are interested here in a different boundary-value problem. In particular, we want to understand how the BH responds when acted upon by an external static perturbation. We will thus start by setting $\tilde \Omega$ to zero in \eqref{RW_conformal}.\footnote{In reality, one might want to set to zero  the frequency $\Omega_*$ of the physical perturbation (see Eq.~\eqref{OmegaStar}).  However, since $\Tilde{\Omega}$ only  enters quadratically  in the master equation \eqref{RW_conformal} and we are interested in the  linear order in $M'$,  whichever frequency is set to zero among $\Omega_*$, $\Tilde{\Omega}$ or ${\Omega}$ is totally immaterial, as this will affect the result only  at order $\mathcal{O}(M'^2)$. }
Note that, as opposed to the case of  an asymptotically-flat Schwarzschild spacetime with  $M'=0$ \cite{Fang:2005qq,Damour:2009vw,Binnington:2009bb,Kol_2012,Hui:2020xxx},  Eq.~\eqref{RW_conformal} has a different singularity structure. This results  in  a change in the form of the falloff of the solutions at the asymptotic boundary. In particular, at large values of $x$, $R(x)$ will not be a simple polynomial. In order to extract the response, we will thus first solve the equation in a region that is sufficiently far from the ``cosmological horizon'', where the standard 
$ x^{\ell}$ and $ x^{-\ell-1}$ falloffs hold. Then,  we will perform a matching with the point-particle effective theory (see Sec.~\ref{ppEFT}), where the response coefficients are defined in a way that is independent of the coordinates.

Concretely, let us start by recalling the expression of the function $f(x)$, and of its derivative, appearing in  Eq.~\eqref{RW_conformal}: 

\begin{equation}
\begin{split}
    f(x)&=4\epsilon|M'|x_C\left(1-\frac{x_H}{x}\right)\left(1-\frac{x}{x_C}\right)\,, \\
    f'(x)&=\frac{4\epsilon|M'|x_H x_C}{x^2}\left(1-\frac{x^2}{x_H\,x_C}\right)\,,
    \label{eq:ffp}
\end{split}
\end{equation}
where $\epsilon$ is again a book-keeping parameter to emphasize that we are taking $ |M'|$ small.
In order to neglect corrections from the singular point at $x_C$, we will focus on the region $x_H \leq x \ll \sqrt{x_H\,x_C}\ll x_C$. This defines a ``near zone'' where both terms $x/x_C$ and $x^2/(x_H \, x_C)$ in \eqref{eq:ffp} are small, and can thus be neglected.\footnote{We stress that, despite the name, the near zone covers a wide region, and is not restricted to values of $x$ close to $x_H$.}   
Introducing  the variable $z\equiv x/x_H$, in the near-zone approximation  Eq.~\eqref{RW_conformal} then reduces to
\begin{multline}
    4 x_C x_H\epsilon|M'| \left[\left(1-\frac{1}{z}\right)R''(z)+\frac{1}{z^2}R'(z)\right]\\
    -\left[\frac{\ell(\ell+1)x_H}{z^2} +\frac{\sigma}{z^3}\right]R(z)=0\,.
\label{LoveEq}
\end{multline}
Note that \eqref{LoveEq}  recovers  the usual master
equation on Schwarzschild spacetime for $\epsilon \rightarrow 0$, as $z\rightarrow x$, $x_H\rightarrow 1$ and $4\epsilon|M'| x_C\rightarrow 1$. It is worth emphasizing that, even though $x_C$ no longer appears as a singular point of  \eqref{LoveEq}, information about $M'$ is still contained in $x_H$.

Eq.~(\ref{LoveEq}) can be recast in a more convenient form by employing the following change of variable and  field redefinition: 
\begin{equation}
\begin{split}
    z&\rightarrow y \equiv \frac{1}{z}\,, \\
    R(y)&\rightarrow p(y)\equiv y^{-\lambda}R(y)\,, \\
\end{split}
\end{equation}
with
\begin{equation}
\begin{split}
    \lambda &=\sqrt{\ell(\ell+1)\,x_H+\frac{1}{4}}-\frac{1}{2} = \\
            &=\ell+4\,\frac{\ell(\ell+1)}{2 \ell +1} \,|M'|\,\epsilon+\mathcal{O}(\epsilon^2)\,.
\end{split}
\end{equation}
With this transformation, one  recovers the canonical form of the hypergeometric equation \cite{seaborn2013hypergeometric,wang1989special}, i.e.,
\begin{equation}
    y(1-y)p''(y)+[c-(a+b+1)y]p'(y)-ab\,p(y)=0\,,
    \label{HypergeometricEquation}
\end{equation}
with $a = \lambda+1-\sqrt{1-\sigma}$, $b = \lambda+1+\sqrt{1-\sigma}$ and $c = 2\lambda+2$. This equation has two linearly independent solutions, which can be expressed as
\begin{equation}
\begin{split}
    &p_1(y)=\leftindex_2 {F}_1\left(a,b;c;y\right)\,, \\
    &p_2(y)=y^{1-c}\leftindex_2 {F}_1\left(a-c+1,b-c+1;2-c;y\right)\,.
\end{split}
\end{equation}
Both $p_1$ and $p_2$ exhibit a logarithmic divergence at the horizon $z=1$. To obtain the physical solution that is regular at the BH horizon,  we  take the  linear combination\footnote{Note that the master variables might not be directly observables. To impose the correct boundary condition at the horizon, one should require that physical observable quantities are regular at $x_H$. 
This can be done by looking for instance at scalar quantities constructed with the master variable solutions. See e.g.~\cite{Hui:2020xxx,Hui:2021vcv}.} 
\begin{equation}
    p(y) = \frac{\Gamma(2-c)}{\Gamma(a-c+1)\Gamma(b-c+1)}\,p_1(y)-\frac{\Gamma(a+b)}{\Gamma(a)\Gamma(b)}\,p_2(y).
\end{equation}
Expanding for $z\gg 1$, one obtains 
\begin{equation}
    R(z)\sim z^{\lambda +1}\Big[1+k \,z^{-2 \lambda +1}\Big]\,,
    \label{solution_z}
\end{equation}
up to an overall constant which corresponds to the amplitude of the external probe tidal field.  The relative coefficient $k$ is 
\begin{equation}
\begin{split}
    k &= \frac{\Gamma(2-c)\Gamma(a)\Gamma(b)}{\Gamma(a-c+1)\Gamma(b-c+1)\Gamma(a+b)}  \\
    &=(a+b-1)\frac{\Gamma(a)^2\Gamma(b)^2}{\Gamma(a+b)^2}\frac{\sin(\pi a) \sin(\pi b)}{\pi\sin(\pi (a+b))}\,,
\end{split}
\label{eq:kcoeffs}
\end{equation}
and  is related to the tidal Love numbers of the BH, as we will see more explicitly below.
Note that  $k$ vanishes for $M' = 0$, which is consistent with the   result that Schwarzschild BHs have vanishing static tidal response. 

Let us focus for a moment on the growing term in Eq.~(\ref{solution_z}). Transforming back to physical coordinates, it reads 
\begin{equation}
    z^{\lambda +1}= \left(\frac{r}{2 M_0}\right)^{\ell+1}(1+\delta_{\text{tf}}(w,r))\,,
\end{equation}
where
\begin{equation}
\begin{split}
    \delta_{\text{tf}}(w,r)\equiv &-(\ell+1)\Bigg[\left(4+s\, \frac{w-w_0}{2 M_0}\right)\\
    &-\frac{4 \ell}{2\ell +1}\,\ln\left(\frac{r}{2 M_0}\right)\Bigg]\,|M'|\,\epsilon+\mathcal{O}(\epsilon^2)\,.
    \label{tidal_correction}
\end{split}
\end{equation}
The function $\delta_{\text{tf}}$ is a correction to the asymptotic profile of the tidal  field induced  by the adiabatic evolution of the BH. 
Note that $\delta_{\text{tf}}$ is a sum of  two terms, a piece that is linear in the null coordinate and a logarithmic term, which both vanish in the $\vert M'\vert \rightarrow 0$ limit.

Plugging the values of $a$ and $b$ for $\sigma = 1,0,-3$ in \eqref{eq:kcoeffs}, we obtain the response coefficients for the various spins, perturbatively in the mass derivative, i.e.,
\begin{equation}
    k = k^{(0)}+k^{(1)}\epsilon+\dots\,,
\end{equation}
where $k^{(0)}=0$ \cite{Fang:2005qq,Damour:2009vw,Binnington:2009bb,Gurlebeck:2015xpa,Hui:2020xxx,Kol_2012},
and the leading-order corrections to the Schwarzschild result for different  spins read
\begin{equation}
\begin{split}
    &k_{\text{scalar}}^{(1)} = \frac{\ell^3(\ell+1)}{2 (2 \ell+1)^2}\frac{\Gamma(\ell)^4}{\Gamma(2 \ell)^2}\,|M'|\,,\\
    &k_{\text{e.m.}}^{(1)} = \frac{\ell(\ell+1)^3}{2 (2 \ell+1)^2}\frac{\Gamma(\ell)^4}{\Gamma(2 \ell)^2}\,|M'|\,,\\
    &k_{\text{grav.~axial}}^{(1)} = \frac{\ell(\ell+2)^2(\ell+1)^3}{2(2\ell+1)^2(\ell-1)^2}\frac{\Gamma(\ell)^4}{\Gamma(2 \ell)^2}\,|M'|\,,
\end{split}
\label{eq:ks}
\end{equation}
where we used the trigonometric identities and the property of the Euler's gamma function, $\Gamma(\ell+1)=\ell\,\Gamma(\ell)$. 

Note that, at linear order in the mass derivative, the Love numbers are constant, while radial running and time evolution appear at quadratic order. 

In summary, at large distances from the BH horizon, but still restricting  to the near-zone regime defined above, the solution to the static equation for a perturbation of spin $\sigma$ is
\begin{equation}
\begin{split}
     R_\sigma&(w,r)\sim \left(\frac{r}{2 M_0}\right)^{\ell+1}\Bigg[1+\delta_{\text{tf}}(w,r)+\mathcal{O}\left(\frac{2 M_0}{r}\right)\\
     &+k_\sigma\,\left(\frac{r}{2 M_0}\right)^{-(2 \ell+1)}\left(1+\mathcal{O}\left(\frac{2 M_0}{r}\right)\right)\Bigg]\,,
\end{split}
\label{fullStaticSolution}
\end{equation}
where $k_\sigma$ can be read off from Eq.~\eqref{eq:ks}.

\subsection{Point-particle EFT}
\label{ppEFT}
In order to gain insight on the physical meaning of the coefficients $k_\sigma$ computed above, we will now resort to the point-particle Effective Field Theory (EFT) \cite{Goldberger:2022rqf,Goldberger:2005cd} (see \cite{Rothstein:2014sra,Porto:2016pyg,Levi:2018nxp,Goldberger:2022ebt} for some reviews). The EFT has the advantage that it defines the induced response in a way that is coordinate independent and more directly related to observable quantities. We will thus introduce the Love numbers as coupling constants of operators localized on the point particle's worldline in the EFT, and compute them by matching with the full solution \eqref{fullStaticSolution}. Again, we will work perturbatively in $M'$, i.e.~we will treat the Vaidya solution as a small correction to the leading Schwarzschild geometry.  
We will discuss in  detail the scalar-field case, and report the result for spin-1 and spin-2 Love numbers at the end.

Indicating with $X^\mu$ a generic set of spacetime coordinates, the point-particle EFT, including finite-size operators, has the following form:
\begin{equation}
S_{\text{EFT}}=S_{\text{bulk}}+S_{\text{scalar}}+S_{\text{pp}}+S_{\text{int}}\,,
\label{EFTaction}
\end{equation}
where $S_{\text{bulk}}$ is the gravitational bulk action  given by
\begin{equation}
    S_{\text{bulk}}=\frac{1}{16\pi}\int {\rm d}^4 X\,\sqrt{-g}\,g^{\mu\nu}\left(R_{\mu\nu} -T_{\mu\nu}\right)\,,
\end{equation}
including the Vaidya SET, $S_{\text{scalar}}$ captures the scalar dynamics in the bulk geometry,
\begin{equation}
    S_{\text{scalar}}=-\frac{1}{2}\int {\rm d}^4 X\,\sqrt{-g}\, g^{\mu\nu}\partial_\mu \Phi\partial_\nu \Phi\,,
\end{equation}
$S_{\text{pp}}$ describes the motion of the point particle, which to leading-order in the small-$M'$ expansion is simply the worldline's Nambu--Goto action on flat space,

\begin{equation}
    S_{\text{pp}}=-M_0\int {\rm d}\tau \, ,
\label{pp_action}
\end{equation}
where $\tau$ parametrizes the worldline, and $S_{\text{int}}$ describes finite-size effects i.e., schematically,
\begin{equation}
    S_{\text{int}}=\int {\rm d} \tau \, e \sum_{\ell=1}^\infty \frac{\mu_\ell}{2\ell!}\left( P^{\nu_1}_{(\mu_1}\cdots P^{\nu_\ell}_{\mu_\ell)_T} \nabla_{\nu_1}\cdots \nabla_{\nu_\ell}\Phi\right)^2 
\end{equation}
where $P^{\nu}_{\mu}$ projects onto the point-particle rest frame, $e$ is an einbein enforcing reparametrization invariance of the worldline, and $(\cdots)_T$ denotes the symmetrized traceless component of the enclosed indices  (see Refs.~\cite{Goldberger:2005cd,Hui:2020xxx,Kol_2012,Goldberger:2022rqf} for details).
The couplings $\mu_\ell$ in  $S_{\text{int}}$ represent the Love numbers, which we want to match with the $k_\sigma$ coefficients derived above.
Note that the EFT action \eqref{EFTaction} contains Vaidya corrections; however, as we shall see below, it will be enough for the matching to solve the equations to leading order, effectively setting $M'$ to zero.

Let us start by expanding the scalar $\Phi$ as $\Phi = \Phi_{\text{tidal}}+\Phi_{\text{resp}}$, where $\Phi_{\text{tidal}}$ represents the external tidal field, which solves the free bulk equation of motion $\Box \Phi_{\text{tidal}}=0$, while $\Phi_{\text{resp}}$ encodes the response that we want to compute and which solves the inhomogeneous equation
\begin{multline}
\Box \Phi_{\text{resp}}  = (-1)^{\ell+1} \frac{\mu_\ell}{\ell!} 
\nabla^{\rho_1}\cdots \nabla^{\rho_\ell}
\Big( \delta_D^{(3)}\left(X-X(\tau)\right)
\\
 \times
P^{(\mu_1}_{\rho_1}\cdots P^{\mu_\ell)_T}_{\rho_\ell}
P^{\nu_1}_{(\mu_1}\cdots P^{\nu_\ell}_{\mu_\ell)_T} \nabla_{\nu_1}\cdots \nabla_{\nu_\ell}\Phi_{\text{tidal}}\Big).
\label{eq:scalareft}
\end{multline}
The idea is to solve \eqref{eq:scalareft} perturbatively in $M'$. Concretely, we shall  expand  each component  in powers of $\epsilon$ as $\Phi_{\text{tidal}}=\Phi_{\text{tidal}}^{(0)}+\epsilon \Phi_{\text{tidal}}^{(1)}+\dots$,   $\Phi_{\text{resp}}=\Phi_{\text{resp}}^{(0)}+\epsilon \Phi_{\text{resp}}^{(1)}+\dots$, and similarly for the Love number couplings, $\mu_\ell=\mu_\ell^{(0)}+\epsilon \mu_\ell^{(1)}+\dots$, and the covariant derivatives. 
However, since we already know that the induced static response  of Schwarzschild black holes is zero in GR~\cite{Fang:2005qq,Damour:2009vw,Binnington:2009bb,Gurlebeck:2015xpa,Hui:2020xxx,Kol_2012}, we can set $\Phi_{\text{resp}}^{(0)}=0=\mu_\ell^{(0)}$, and just focus on linear quantities in $\epsilon$. The vanishing of  $\mu_\ell^{(0)}$ implies that, at linear order in $\epsilon$,  we can replace $\Phi_{\text{tidal}}$   on the right-hand side of \eqref{eq:scalareft} with $\Phi_{\text{tidal}}^{(0)}$ and replace the covariant derivatives with simple derivatives in Minkowski space.
The zeroth-order tidal field can be written, in cartesian coordinates, as 
\begin{equation}
\Phi_{\text{tidal}}^{(0)}=c_{a_1\dots a_\ell}X^{a_1}\dots X^{a_\ell}\,,
\end{equation}
or, equivalently, in spherical harmonics as $C_\ell r^\ell Y_{\ell m}(\theta,\phi)$ with some overall $\ell$-dependent amplitude coefficient \cite{Hui:2020xxx}. Similarly, since $\Phi_{\text{resp}}^{(0)}=0$,  on the left-hand side of \eqref{eq:scalareft}, we can replace $\Phi_{\text{resp}}$ with $\Phi_{\text{resp}}^{(1)}$ and the covariant d'Alembert operator with the Laplace operator in flat space.
All in all, the response field equation boils down to 
\begin{equation}
    \vec{\nabla}^2 \Phi_{\text{resp}}^{(1)}= J \,,
    \label{nonhom_KG}
\end{equation}
with the source term on the right-hand side given, in the rest frame of the point particle, by
\begin{equation}
    J = \mu_\ell(-1)^{\ell+1}\,c^{a_1\dots a_\ell}\partial_{(a_1}\dots \partial_{a_\ell)_T}\delta_D^{(3)}(\vec X)\,.
\end{equation}
Eq.~(\ref{nonhom_KG}) can be solved in Fourier space. From the  Green's function of the Laplace operator in flat space, $G(\Vec{p})=-|\Vec{p}|^{-2}$, one obtains \cite{Hui:2020xxx}
\begin{equation}
    \Tilde{\Phi}_{\text{resp}}^{(1)}(\vec{p})=\mu_\ell(-i)^\ell\,c^{a_1\dots a_\ell}\frac{p_{(a_1}\dots p_{a_\ell)_T}}{|\Vec{p}|^2}\,.
\end{equation}

Finally, from the inverse Fourier transform, 
\begin{equation}
\begin{split}
   \Phi_{\text{resp}}^{(1)} & =\mu_\ell (-i)^\ell \int\frac{ {\rm d}^3\vec{p}}{(2\pi)^3} \,{\rm e}^{i \vec{p} \cdot \vec{X}} c^{a_1 \dots a_\ell} \frac{ p_{(a_1} \dots p_{a_\ell)_T} }{|\vec{p}|^2}   \\
    &=\mu_\ell K\,
c_{a_1 \cdots a_\ell} X^{a_1} \cdots X^{a_\ell} \,\left(\frac{| \vec{X}|^2}{4} \right)^{- \frac{1}{2} - \ell}\,,
\end{split}
\end{equation}
with the prefactor
\begin{equation}
    K = \frac{(-1)^\ell}{2^{\ell+3}\sqrt{\pi} \Gamma\left(\frac{1}{2}-\ell\right)}\,.
    \label{Matching_Spin0}
\end{equation}
By comparing the EFT solution $\Phi_{\text{resp}}^{(1)}$ with the full solution $\Phi=  R(x)Y_{\ell m}(\theta,\phi)/x$ where $R(x)$ can be read off from \eqref{fullStaticSolution}, to linear order in $\epsilon$,   
we can express the EFT Love number coefficients $\mu_\ell$ in terms of $k_{\text{scalar}}^{(1)}$ in \eqref{eq:ks}. The result is
\begin{equation}
    \mu_\ell^{(1)}= \frac{(-1)^\ell \sqrt{\pi} }{2^{\ell-2}}\Gamma\left(\frac{1}{2}-\ell\right)\left(2 M_0\right)^{2\ell+1}\,k_{\text{scalar}}^{(1)}\,.
\end{equation}

A similar procedure can be followed in the case of electromagnetic and gravitational axial perturbations.  To describe electromagnetic response, 
one has to introduce in the bulk  the electromagnetic action
\begin{equation}
    S_{\rm em}=-\frac{1}{4}\int {\rm d}^4x\,\sqrt{-g}F_{\mu\nu}F^{\mu\nu}\,.
\end{equation}
The metric is expanded in perturbations as
\begin{equation}
    g_{\mu\nu}=\g^{(B)}_{\mu\nu}+h_{\mu\nu}\,,
\end{equation}
where $g^{(B)}_{\mu\nu}$ is the Minkowski flat metric plus perturbative corrections in $M'$.  The interacting part of the action can be constructed as
\begin{equation}
\begin{split}
     S_{\text{int}}&=\int {\rm d} \tau \, e \sum_{\ell} \Bigg[\frac{\mu_\ell^E}{2\ell!}\left( P^{\nu_1}_{(\mu_1}\cdots P^{\nu_\ell}_{\mu_\ell)_T} \nabla_{\nu_1}\cdots \nabla_{\nu_{\ell-1}}E_{\nu_{\ell}}\right)^2 \\
     &+\frac{\mu_\ell^B}{4\ell!}\left( P^{\nu_1}_{(\mu_1}\cdots P^{\nu_\ell}_{\mu_\ell)_T} P^{\beta}_{\alpha}\nabla_{\nu_1}\cdots \nabla_{\nu_{\ell-1}}B_{\nu_{\ell} \beta}\right)^2\\
     +&\frac{\mu_\ell^{C_B}}{4\ell!}\left( P^{\nu_1}_{(\mu_1}\cdots P^{\nu_\ell}_{\mu_\ell)_T} P^{\beta}_{\alpha}\nabla_{\nu_1}\cdots \nabla_{\nu_{\ell-2}}B^{(2)}_{\nu_{\ell-1}\nu_{\ell} |\beta}\right)^2\Bigg]\,,
\end{split}
\end{equation}
where the sum over $\ell$ starts from $\ell = 1$ for the first two terms and from $\ell = 2$ for the last one. The objects $E_\mu$, $B_{\mu\nu}$ e $B^{(2)}_{\mu\nu\beta}$ are related  to the electric and magnetic fields, and the magnetic part of the Weyl tensor, respectively. In the rest frame of the point particle, their only nonzero components, at leading order in the flat-space limit, are
\begin{equation}
\begin{split}
    &E_i = 
    -\partial_i A_t\, , \\
    & 
    B_{ij}=\partial_i A_j-\partial_j A_i\,  , \\
    & 
    B^{(2)}_{ij|k}=C_{0ijk}\,,
\end{split}
\end{equation}
where $A_\mu$  is the electromagnetic potential and $C_{\mu\nu\alpha\beta}$ is the Weyl tensor. This is all that we will need on the EFT side.

As already discussed for the scalar case, one can split the field $A_\mu$ and $h_{\mu\nu}$ in a tidal part and a response part, and then expand both to linear order in $\epsilon$. Since the zeroth-order response is vanishing, one obtains an inhomogeneous equation for the linear correction in the form \eqref{nonhom_KG}, with a source term involving derivative computed on a flat background. The relevant equations for linear electromagnetic and gravitational response can be found in \cite{Rai:2024lho,Hui:2020xxx}. 
To perform the matching, we shall use  
the $t$-component of the four-potential, the angular components of  $B_{ij}$ and the components $C_{trij}$ of the Weyl tensor. The latter can be directly related to the Regge--Wheeler master variable $\psi$ defined as auxiliary field in Appendix \ref{RW_Equation} in the static limit.

The result of the matching is:
\begin{equation}
\begin{split}
    &\mu_\ell^{E,(1)}=\frac{(-1)^{\ell+1}\ell \sqrt{\pi}}{ 2^{\ell-1}(\ell+1)}\Gamma \left(\frac{1}{2}-\ell\right)\left(2 M_0\right)^{2\ell+1}k_{\rm e.m.}^{(1)} \,,\\
    &\mu_\ell^{B,(1)}=\frac{(-1)^\ell \ell\sqrt{\pi}}{ 2^{\ell-1}(\ell+1)}\Gamma \left(\frac{1}{2}-\ell\right)\left(2 M_0\right)^{2\ell+1}k_{\rm e.m.}^{(1)} \,,\\
    &\mu_\ell^{C_B,(1)}=\frac{(-1)^{\ell+1}\ell(\ell-1) }{\sqrt{\pi} (\ell+1)(\ell+2)2^{\ell+1}}\Gamma \left(\frac{1}{2}-\ell\right)\times \\
    &\qquad\qquad\qquad\qquad\qquad\times\left(2 M_0\right)^{2\ell+1}k_{\rm grav}^{(1)}\,. \\
\end{split}
\label{Matching_Spin12}
\end{equation}

The relations in Eq.(\ref{Matching_Spin0}) and Eq.(\ref{Matching_Spin12}) are analogous to the ones that hold for the Schwarzschild BH. However, this computation demonstrates the robustness of the result of the previous section, namely that the Vaidya BH, unlike Schwarzschild, exhibits a nontrivial tidal response, quantified by the Love numbers presented in Eqs (\ref{eq:ks}).

\section{Conclusions}

In this work, we  studied perturbations of  Vaidya BHs. In the first part, we revisited the computation of the QNMs. Unlike previous works in the same context \cite{Shao:2004ws, Abdalla:2006vb,Chirenti:2010iu, Chirenti:2011rc, Lin:2021fth, Redondo-Yuste:2023ipg}, our analysis was carried out in the frequency domain, under the  assumption  of a constant  rate of change $M'$ of the BH mass.

In this framework, we computed the QNM spectrum both  analytically  in the eikonal approximation, and with a  numerical approach based on Leaver's continued-fraction method. 
One main advantage of our approach is that,
with respect to previous results in the literature, 
it allowed us to compute the frequencies with higher  accuracy. In addition, it provides a more transparent understanding of the physical effects associated with the dynamical BH mass, including the stability properties of the solution.

On the other hand, the fact that the (null) time evolution of the physical frequencies inversely tracks the dynamical BH mass, which was found numerically in \cite{Abdalla:2006vb}, is naturally recovered in our formalism, through the scaling of the physical frequencies $\omega\propto M(w)^{-1}$, as discussed in Sec. \ref{time_domain}.

In the second part, we studied tidal effects on Vaidya BHs. As opposed to the QNM analysis, we worked here under the assumption that $M'$ is small  (in addition to being constant), and we computed explicitly the tidal Love numbers to linear order in $M'$. Our result shows that dynamical effects due to a time-dependent mass of the Vaidya BH induce non-vanishing tidal response. This should be contrasted with the case of Schwarzschild BHs, whose Love numbers are exactly zero in GR.

\section*{Acknowledgements}
We thank Nicola Franchini for insightful discussions.

E.B. and L.C. acknowledge support from the European Union’s H2020 ERC Consolidator Grant ``GRavity from Astrophysical to Microscopic Scales'' (Grant No. GRAMS-815673), the PRIN 2022 grant ``GUVIRP - Gravity tests in the UltraViolet and InfraRed with Pulsar timing'', and the EU Horizon 2020 Research and Innovation Programme under the Marie Sklodowska-Curie Grant Agreement No. 101007855.
The work of L.S.~was supported by the Programme National GRAM of CNRS/INSU with INP and IN2P3 co-funded by CNES.

\appendix
\section{Perturbation equation in Eddington-Finkelstein coordinates}
\label{perturbation_eqs}
In the following, we  derive the equations for perturbations of various spin 
in the Jordan frame and in  the Eddington--Finkelstein coordinates. We will mainly work at the level of the action, and show that the result recovers  the equations obtained in the main text, after transforming into rescaled coordinates. For the sake of the presentation, we just illustrate the decreasing mass case (where $w$ is the retarded time), as the increasing mass case is completely analogous. 
\label{equation derivation}

\subsection{Scalars and photons}
The dynamics of scalar perturbations obeys the Klein--Gordon equation 
\begin{equation}
    \partial_\mu\left(\sqrt{-g}\,g^{\mu\nu}\partial_\nu \Phi \right)=0\,.
\label{KG}
\end{equation}
The scalar field can be expanded as
\begin{equation}
    \Phi(w,r,\theta,\phi)=\sum_{\ell,m}\frac{f(w,r)}{r}Y_{\ell m}(\theta,\phi)\,.
    \label{KG_master_variable}
\end{equation}

With this ansatz, on the Vaidya background 

Eq.~(\ref{KG}) reads:
\begin{multline}
   \left(1-\frac{2 M(w)}{r}\right)\frac{\partial^2 f(w,r)}{\partial r^2}-2\frac{\partial^2 f(w,r)}{\partial w\partial r}\\
    +\frac{2\,M(w)}{r^2}\frac{\partial f(w,r)}{\partial r}\\
    -\left(\frac{\ell(\ell+1)}{r^2}+\frac{2  M(w)}{r^3}\right)f(w,r)= 0\,.
\label{KG_Vaidya} 
\end{multline}

Note that this equation is equivalent to the one derived in the Einstein frame in Sec.~\ref{scalar_equation}. In fact, the master variable of Eq.~(\ref{KG_master_variable}) implicitly introduces a conformal factor $(2 M(w))^{-1}$ encoded in the term $r^{-1}$. This corresponds to the conformal weight $\chi = -1$ that we choose for the scalar field in the main text.

Electromagnetic perturbations are described by the Maxwell action
\begin{equation}
    S_{\text{EM}}[A,g]=-\frac{1}{4}\int {\rm d}^4x\,\sqrt{-g}\, F_{\mu\nu}F^{\mu\nu}\,,
\label{Maxwell_action}
\end{equation}
with the field strength $F_{\mu\nu}=\partial_{[\mu}A_{\nu]}$. 
The four-potential in the coordinates $(w,r,\theta,\phi)$ can be decomposed with the ansatz
\begin{equation}
\begin{split}
        A &= A^{\text{(polar)}}+A^{\text{(axial)}}=\\
        &=\left(\begin{array}{cccc}
 h(w,r)\,Y_{\ell m}\\
 e(w,r) \,Y_{\ell m}\\
 a(w,r) {\varepsilon_i}^j\,\partial_jY_{\ell m}+k(w,r)\,\partial_iY_{\ell m}
\end{array}
\right) \,.
\label{Four_potential}
\end{split}
\end{equation}
Note that the  $\ell, m$ dependence has been left implicit in the functions $h$, $e$, $k$ and $a$. Adopting the gauge choice $k(w,r)=0$, and after integrating out the angular part, one obtains the reduced action
\begin{equation}
    S_{\text{red}}=\int {\rm d}w\,{\rm d}r\,\mathcal{L}[a,\,h,\,e\,]\,,
\end{equation}
with
\begin{equation}
\begin{split}
    \mathcal{L}[a,\,h,\,e\,]=& \, \frac{\ell(\ell+1) }{2 } \Bigg[-\ell(\ell+1) \frac{a^2}{r^2}+2 \partial_u a  \partial_r a+\\
    &-\left(1-\frac{2 M(w)}{r}\right) \partial_ra^2+\\
    &-\left(1-\frac{2 M(w)}{r}\right) e^2+\\
    &+\frac{r^2 \left( \partial_r h- \partial_u e\right)^2}{\ell(\ell+1)}+2  e\, h \Bigg]=\\
    =&\,\mathcal{L}[a]+\mathcal{L}[h,\,e]\,,
\end{split}
\end{equation}
where we defined the Lagrangians $\mathcal{L}[a]$ and $\mathcal{L}[h,\,e]$   describing  the axial degree of freedom $a$, and the polar modes $(h,\,e)$, respectively. 

Varying $\mathcal{L}[a]$ with respect to $a$, one gets the equation of motion
\begin{equation}
    \begin{split}
   \Bigg(1-\frac{2 M(w)}{r}&\Bigg)\frac{\partial^2 a(w,r)}{\partial r^2}-2\frac{\partial^2 a(w,r)}{\partial w\partial r}+\\
    +&\frac{2\,M(w)}{r^2}\frac{\partial a(w,r)}{\partial r}+\\
    &-\frac{\ell(\ell+1)}{r^2}a(w,r)= 0\,.
\end{split}
\label{a_eom}
\end{equation}

Instead, $\mathcal{L}[h,\,e]$ can be conveniently rewritten, introducing an auxiliary field $q$, as
\begin{equation}
\begin{split}
    \mathcal{L}[h,\,e,\,q]=&\frac{\ell(\ell+1)}{2} \Bigg[-\frac{\ell(\ell+1) q^2}{2 r^2}+\\
    &+q\,(\partial_r h-\partial_w e)+2e h+\\
    &\qquad\qquad+\left(\frac{2M(w)}{r}-1\right) e^2\Bigg].
\end{split}
\label{eq:actionq}
\end{equation}
It is straightforward to check that \eqref{eq:actionq} is equivalent to $\mathcal{L}[h,\,e]$  upon using the $q$'s equation of motion,
\begin{equation}
    q(w,r) = \frac{r^2}{\ell(\ell+1)}\left(\partial_r\,h(w,r)-\partial_u e(w,r)\right)\,.
    \label{q_master_variable}
\end{equation}

The introduction of the field $q$ is convenient because it makes it easier  to integrate out the fields $e$ and $h$ from \eqref{eq:actionq}. Computing the equations of motion of $e$ and $h$, and plugging the solutions back into \eqref{eq:actionq}, one finds the following action for the single degree of freedom $q$:
\begin{equation}
\begin{split}
    \mathcal{L}[q]=&-\frac{\ell(\ell+1)}{2 }  \Bigg[\ell(\ell+1) \frac{q^2}{r^2}+\\
    &+\left(1-\frac{2 M(w)}{r}\right) \partial_r q^2-2  \partial_w q\partial_rq\Bigg]\,,
\end{split}
\end{equation}
yielding the equation of motion 
\begin{equation}
     \begin{split}
   \Bigg(1-\frac{2 M(w)}{r}&\Bigg)\frac{\partial^2 q(w,r)}{\partial r^2}-2\frac{\partial^2 q(w,r)}{\partial w\partial r}+\\
    +&\frac{2\,M(w)}{r^2}\frac{\partial f(w,r)}{\partial r}+\\
    &-\frac{\ell(\ell+1)}{r^2}q(w,r)= 0\,.
\end{split}
\end{equation}
Note that this has the same for as Eq.~(\ref{a_eom}). This fact, which is responsible for isospectrality of the even and odd electromagnetic modes, is a consequence of the electric-magnetic duality, as we will now explicitly show, mirroring exactly the Schwarzschild case \cite{Hui:2020xxx}.

Notice also that, like in the scalar case, we are implicitly introducing a conformal factor through the master variable of Eq. (\ref{q_master_variable}). The conformal weight is related to the power of $r$ that appears in the definition on $q(w,r)$. Indeed, $q\sim (2 M(w)) \,h$, and, since $F\sim \partial q$ we have the implicit transformation law for the field strength $F\rightarrow F$. This corresponds to the choice of $\chi = 0$ that we made in the Einstein frame in Sec. \ref{em_equations}.

\subsection{Electric-magnetic duality}
The two electromagnetic degrees of freedom $a(w,r)$ and $q(w,r)$ can be joined in the $SO(2)$ doublet
\begin{equation}
    \xi(w,r) = \left(\begin{array}{cccc}
 a(w,r) \\
 q(w,r)
\end{array}\right)\,,
\end{equation}
whose dynamics is described by the action
\begin{equation}
\begin{split}
    S[\xi]=&-\frac{\ell(\ell+1)}{2}\int{\rm d}w\,{\rm d}r\Bigg[ \frac{\ell(\ell+1)}{r^2}\xi^T\,\xi+ \\
    &+\left(1-\frac{2 M(w)}{r}\right) \partial_r \xi^T\,\partial_r\xi+\\
    &\qquad\qquad-2 \,\partial_u \xi^T\,\partial_r\xi\Bigg]\,.
\end{split}
\end{equation}
The Maxwell equations can be written concisely using differential forms as
\begin{equation}
    {\rm d}F = 0 \, , \qquad
    {\rm d}\star F = 0 \, ,
\end{equation}
where $\star$ represents the Hodge dual operation. This form makes it evident that the equations are symmetric under $F \leftrightarrow \star F$, which is the well-known electric-magnetic duality.  

The components of the field strength can be computed explicitly from the four-potential in Eq.~(\ref{four_potential})
\begin{equation}
\begin{split}
    &F_{wr}=-\frac{\ell(\ell+1)}{2r^2}q\,Y_{\ell m}\\
    &F_{wi}=\partial_w a\,{\varepsilon_i}^j\,\partial_jY_{\ell m}+\\
    &\qquad\qquad-\Bigg[\Bigg(1-\frac{2 M(w)}{r}\Bigg)\,\partial_r q-\partial_wq\Bigg]\,\partial_iY_{\ell m}\\
    &F_{ri}=\partial_r a\,{\varepsilon_i}^j\,\partial_jY_{\ell m}-\partial_r q,\partial_iY_{\ell m}\\
    &F_{ij}=-\frac{\ell(\ell+1)}{2}\epsilon_{ij}\,a\,Y_{\ell m}\,,
\end{split}
\end{equation}
where the indices $i,j$ run over the angular coordinates. One can compute the dual components $\star F_{\alpha\beta}=\frac{1}{2}\epsilon_{\alpha\beta\mu\nu}F^{\mu\nu}$ and verify that they yield precisely the  same expressions, upon exchanging $a\leftrightarrow q$.

\subsection{Regge--Wheeler equation}
\label{RW_Equation}
The gravitational perturbations, like the electromagnetic ones, can be decomposed into axial and polar sectors \cite{PhysRev.108.1063}. 
Such decomposition is useful as the spherical symmetry of the Vaidya geometry ensures that the two sectors are decoupled at the level of the linearized dynamics. In the following, we will focus on the axial sector only.

The axial metric perturbations can be expressed, in the coordinates $(w,r,\theta,\phi)$, and in the Regge--Wheeler gauge~\cite{PhysRev.108.1063}, as
\begin{equation}
\begin{split}
h_{\mu\nu}^{(\text{axial})}=&
\sum_{\ell m}
\left(
\begin{array}{cccc}
 0 & 0 & -\frac{h_0}{\sin \theta}\, \partial_\phi  & \sin \theta \,h_0 \, \partial_\theta  \\
 0 & 0 &  -\frac{h_1}{\sin \theta}\,\partial_\phi  & \sin \theta \,h_1 \, \partial_\theta \\
\text{Sym} & \text{Sym} & 0 & 0 \\
\text{Sym} & \text{Sym} & 0 & 0 \\
\end{array}
\right)\\
&\times Y_{\ell m}(\theta ,\phi )\,,
\end{split}
\label{TransformedPerturbation}
\end{equation}
where $h_0$ and $h_1$ are functions of $w$ and $r$.
The dynamics of the odd fields can be obtained from the action

\begin{equation}
S_{\text{grav}}=\frac{M_{\text{Pl}}^2}{2}\int {\rm d}^4x\,\sqrt{-g}\big(R-{T^\lambda}_\lambda\big)\,,
\label{Einstein_Hilbert_Action}
\end{equation}
where ${T^\lambda}_\lambda$  is the trace of the stress-energy tensor of Eq.~(\ref{Vaidya_SET})), sourcing the Vaidya geometry.

Expanding \eqref{Einstein_Hilbert_Action} to quadratic order in the fields, and integrating over the solid angle, one finds an action for $h_0$ and $h_1$ which is again more conveniently rewritten in terms of an auxiliary field $\psi$ as 
\begin{equation}
    S^{(2)}_{\rm axial}
    =-\frac{\,\ell(\ell+1)M_{\text{Pl}}^2}{2}\int {\rm d}w\,{\rm d}r\,\mathcal{L}[h_0,\,h_1,\,\psi]\,,
\end{equation}
where 
\begin{multline}
    \mathcal{L}[h_0,h_1,\psi] = \frac{r^2 \psi^2}{2}+\psi \left(-2 h_0+r(\partial_r h_0-\partial_w h_1)\right) \\
    +\frac{(\ell+2)(\ell-1)}{r^2}\,\left[\left(1-\frac{2 M(w)}{r}\right)\frac{h_1^2}{2}
    -h_0 h_1\right],
\label{eq:actionh0h1app}
\end{multline}
where the auxiliary field $\psi$ is expressed in terms of $h_0$ and $h_1$ via the relation

\begin{equation}
    \psi(w,r)=2h_0(w,r)+r\left(\partial_w h_1(w,r)-\partial_r h_0(w,r)\right)\,.
    \label{psi_definition}
\end{equation}
Integrating out $h_0$ and $h_1$ from \eqref{eq:actionh0h1app}, one finally finds a quadratic action for the single degree of freedom $\psi$:
\begin{equation}
\begin{split}
    \mathcal{L}[\psi]=&-\left(1-\frac{2 M(w)}{r}\right) \frac{\partial_r\psi^2}{2}+\partial_r\psi \partial_w\psi +\\
    &-\left(\frac{\ell(\ell+1)}{r^2}-\frac{6 M(w)}{r^3}\right) \frac{\psi^2}{2} \,,
\end{split}
\end{equation}
which yields the equation of motion
\begin{equation}
    \begin{split}
   \Bigg(1-&\frac{2 M(w)}{r}\Bigg)\frac{\partial^2 \psi(w,r)}{\partial r^2}-2\frac{\partial^2 \psi(w,r)}{\partial w\partial r}+\\
    +&\frac{2\,M(w)}{r^2}\frac{\partial\psi(w,r)}{\partial r}+\\
    &-\left(\frac{\ell(\ell+1)}{r^2}-\frac{6 M(w)}{r^3}\right)\psi(w,r)= 0\,,
\end{split}
\end{equation}
representing the generalization of  the Regge--Wheeler equation to  Vaidya BHs.

\subsection{Master equation and spectrum}
All in all, we can write down a single master equation (holding for both the increasing and decreasing mass cases) for massless scalar, (polar and axial) electromagnetic and (axial) gravitational perturbations on a Vaidya background in Eddington--Finkelstein coordinates \cite{Abdalla_2007,Chirenti:2010iu,Redondo-Yuste:2023ipg}:
\begin{equation}
\begin{split}
   \Bigg(1-&\frac{2 M(w)}{r}\Bigg)\frac{\partial^2 \psi(w,r)}{\partial r^2}+2s\,\frac{\partial^2 \psi(w,r)}{\partial u\partial r}+\\
    +&\frac{2\,M(w)}{r^2}\frac{\partial\psi(w,r)}{\partial r}+\\
    &-\left(\frac{\ell(\ell+1)}{r^2}+\frac{2 \,\sigma M(w)}{r^3}\right)\psi(w,r)= 0\,,
\end{split}
\label{RW_Vaidya}
\end{equation}
where $\sigma = 1,0,-3$ holds respectively for scalar, (polar and axial) electromagnetic  and axial gravitational perturbations.  

Consider now the change of coordinates introduced in Eq.~(\ref{coordinate_transformation}). The derivatives transform as
\begin{equation}
\begin{split}
    &\frac{\partial}{\partial r}\rightarrow \frac{1}{2 M(w)}\frac{\partial}{\partial x} \\
    &\frac{\partial}{\partial w}\rightarrow\frac{1}{2 M(w)} \frac{\partial}{\partial W}-x\frac{M'(w)}{M(w)}\frac{\partial}{\partial x} \\
    &\frac{\partial^2}{\partial r^2}\rightarrow\frac{1}{4M(w)^2}\frac{\partial^2}{\partial x^2} \\
    &\frac{\partial^2}{\partial r\partial w}\rightarrow \frac{1}{4M(w)^2}\frac{\partial^2}{\partial W\partial x}-\frac{M'(w)}{2M(w)^2}\frac{\partial}{\partial x}+\\
    &\qquad\qquad-x\frac{M'(w)}{2M(w)^2}\frac{\partial^2}{\partial x^2}
\end{split}
\end{equation}
Note that the term $\partial M(w)/\partial x$ is vanishing (from the inverse Jacobian of the coordinate transformation). 
Approximating the mass evolution with a linear growth in the null time, we get that,

modulo an overall factor, Eq.~(\ref{RW_Vaidya})  becomes independent of $u$: 
\begin{multline}
    \frac{\left( x-1-4 |M'|\,x^2\right)}{x}\frac{\partial^2\psi(W,x)}{\partial x^2}+ 2s\, \frac{\partial^2\psi(W,x)}{\partial U\partial x}\\
    +\frac{ \left(1-4 |M'|\,x^2 \right)}{x^2}\frac{\partial\psi(W,x)}{\partial x}\\
    -\left(\frac{\ell(\ell+1)}{x^2} +\frac{ \sigma}{x^3} \right) \psi (W,x)=0\,.
\label{time_independent_eq}
\end{multline}

Given the time-independent form of Eq.~(\ref{time_independent_eq}), we can further change coordinate  from  $W$ to  $T= W-s \, x_*$,  and obtain

\begin{equation}
\begin{split}
    &\frac{\partial}{\partial x}\rightarrow - \frac{s}{f(x)}\frac{\partial}{\partial T}+\frac{\partial}{\partial x}\\
    &\frac{\partial}{\partial W}\rightarrow \frac{\partial}{\partial T} \\
    &\frac{\partial^2}{\partial x^2}\rightarrow \frac{1}{f(x)^2}\frac{\partial^2}{\partial T^2}- \frac{2s}{f(x)}\frac{\partial^2}{\partial T\partial x}+ s\,\frac{f'(x)}{f(x)^2}\frac{\partial}{\partial T} \\
    &\qquad\qquad +\frac{\partial^2}{\partial x^2} \\
    &\frac{\partial^2}{\partial W\partial x}\rightarrow- \frac{s}{f(x)}\frac{\partial^2}{\partial T^2}+\frac{\partial^2}{\partial T\partial x}\,.
\end{split}
\end{equation}

Finally,   
we can use separation of variables in $x$ and $T$ 
by writing  the usual Fourier  $\psi(T,x)=\exp{(-i\Omega_\psi T)}\Psi(x)$. Introducing again the generalized tortoise coordinate, we get the very simple  
master equation 
\begin{equation}
    \Bigg[\frac{{\rm d}^2}{{\rm d}x_*^2}+\Big(\Omega_\psi^2-V(x)\Big)\Bigg]\Psi(x_*)=0\,,
\label{odd_equation}
\end{equation}
with 
\begin{equation}
    V(x)=f(x) \left( \frac{\ell(\ell+1)}{x^2}+\frac{\sigma}{x^3} \right)\,.
\label{Vaidya_RW_Potential}
\end{equation}

Note that this equation recovers Eq.~(\ref{RW_conformal}), upon identifying 
$\Tilde{\Omega}=\Omega_\psi$. From the discussion of Sec.~\ref{sec:gravpts}, in the gravitational case, we thus expect the following relation between the frequencies:

\begin{equation}
    \Omega_\psi = \Omega +2i M'\,.
\end{equation}
This relation can be understood as follows.

The master variable of Eq.~(\ref{psi_definition}) has the same spectrum of the component of the metric perturbations in the Jordan frame $h_{\mu\nu}$. On the other hand, the derivation of Eq.~(\ref{RW_conformal}) was performed in the Einstein frame, namely in terms of the metric perturbation $\Tilde{h}\sim h/(2 M(w))^2$. However, one also has to account for an extra $2 M(w)$ factor  relating the off-diagonal metric perturbation components $h_{t j}$, where $t$ is a ``physical'' time coordinate and $j =\theta,\phi$, with the components $h_{T j}$, where $T$ is the rescaled time coordinate, i.e.,

\begin{equation}
    h_{t j}=(2 M(w))^{-1}\, h_{Tj}=2 M(w) \,\Tilde{h}_{T j}\,.
\end{equation}
As we discussed in the main text, a factor $2 M(w)$ provides the shift $2 i M'$ (see Eq. (\ref{OmegaStar})) and hence this relation yields exactly the expected connection between the two spectra.

\section{Coefficients of the five-term recurrence relation}
\label{5t relation}
In this appendix, we report  the full expressions for the coefficients of the initial five-term recurrence relation introduced in the continued-fraction method. As an example, we show explicitly the decreasing mass case, corresponding to the outgoing Vaidya metric, but the increasing mass case works analogously:
\begin{widetext}
\begin{equation}
\begin{split}
    \alpha_n = &-(n+1) \left(16 \left| M'\right| -1\right) \Bigg[2 \left| M'\right| \Big[8 \left| M'\right| \left(-16 (n+1) \left| M'\right| +n \left(3-2 \sqrt{1-16 \left| M'\right| }\right) \right. \\
    & \left.\left.\left.\left. -2 \sqrt{1-16 \left| M'\right| }+4 \rho +3\right) +n \left(\sqrt{1-16 \left| M'\right| }-1\right) +2 \rho \left(3 \sqrt{1-16 \left| M'\right| }-5\right) \right.\right.\right. \\
    &  +\sqrt{1-16 \left| M'\right| }-1\Big] +\rho \left(-\sqrt{1-16 \left| M'\right| }\right) +\rho \Bigg]\,,
\end{split}
\end{equation}

\begin{equation}
\begin{split}
    \beta_n = & \, 2 \left| M'\right|  \Bigg[-4 \left| M'\right|  \Bigg(16 \left| M'\right|  \left(-2 \ell(\ell+1) \left(\sqrt{1-16 \left| M'\right| }-2\right) \right. \\
    & \left.\left.\left. +16 \left(6 n^2+3 n+\sigma \right) \left| M'\right| +(5 n (2 n+1)+\sigma ) \sqrt{1-16 \left| M'\right| } -2 (5 n (2 n+2 \rho +1)+2 \rho +\sigma )\right) \right.\right. \\
    & \left.\left. +\ell(\ell+1) \left(6 \sqrt{1-16 \left| M'\right| }-8\right) -\sqrt{1-16 \left| M'\right| } (3 n (6 n+16 \rho +3)+8 \rho +\sigma ) \right.\right. \\
    & \left. +22 n^2+n (84 \rho +11)+4 \rho  (4 \rho +3)+\sigma \Bigg) +\ell(\ell+1)\left(\sqrt{1-16 \left| M'\right| }-1\right) \right. \\
    & \left. -\left(2 n^2+20 n \rho +n+2 \rho  (6 \rho +1)\right) \sqrt{1-16 \left| M'\right| } +2 n^2+24 n \rho +n+2 \rho  (10 \rho +1)\Bigg] \right. \\
    &  +\rho  (n+2 \rho ) \left(\sqrt{1-16 \left| M'\right| }-1\right) \,,
\end{split}
\end{equation}

\begin{equation}
\begin{split}
    \gamma_n = &2 \left| M'\right|  \Bigg[4 \left| M'\right|  \Bigg(8 \left| M'\right|  \left(2 \ell(\ell+1) \left(5-3 \sqrt{1-16 \left| M'\right| }\right) +16 (13 (n-1) n+5 \sigma ) \left| M'\right|\right. \\
    & \left.\left.\left.  +4 (4 (n-1) n+\sigma ) \sqrt{1-16 \left| M'\right| } +n (-33 n-32 \rho +33)+20 \rho -9 \sigma \right) +\ell(\ell+1) \left(7 \sqrt{1-16 \left| M'\right| }-9\right) \right.\right. \\
    & \left.\left. -2 \sqrt{1-16 \left| M'\right| } (2 n (3 n+7 \rho -3)-9 \rho +\sigma ) +2 \left(20 n \rho +7 (n-1) n+4 \rho ^2-13 \rho +\sigma \right)\Bigg)\right.\right. \\
    &   -\left(\sqrt{1-16 \left| M'\right| }-1\right) \left(\ell(\ell+1)-n(n+6 \rho-1)-4 (\rho -1) \rho \right) \Bigg]\,,
\end{split}
\end{equation}

\begin{equation}
\begin{split}
    \delta_n = 8 & \left| M'\right| ^2 \Bigg[8 \left| M'\right| \Big[2 \ell(\ell+1) \left(\sqrt{1-16 \left| M'\right| }-1\right) -16 (3 n (2 n-5)+4 \sigma +6)\left| M'\right| \\
    & \left.  -2 (n (2 n-5)+\sigma ) \sqrt{1-16 \left| M'\right| } -4 \sqrt{1-16 \left| M'\right| } +n (10 n+8 \rho -25)\right. \\
    & -12 \rho +6 \sigma +10\Big] -\left(\sqrt{1-16 \left| M'\right| }-1\right) \left(\ell(\ell+1)+n (-2 n-4 \rho +5)+6 \rho -\sigma -2\right) \Bigg]\,,
\end{split}
\end{equation}

\begin{equation}
\zeta_n = 64 \epsilon^3 (16 \epsilon +1) ((n-4) n+\sigma +3) .
\end{equation}
\end{widetext}

\clearpage
\bibliographystyle{ieeetr}
\bibliography{Bibliography}

\end{document}